\documentclass[fleqn,usenatbib]{mnras}
\usepackage{array}
\usepackage{subfig}
\usepackage{bm}
\usepackage{newtxtext,newtxmath}
\usepackage[T1]{fontenc}
\usepackage{ae,aecompl}
\usepackage{graphicx}
\usepackage{amsmath}
\usepackage{subfig}
\usepackage{float}
\usepackage{threeparttable}
\usepackage{comment}
\restylefloat{figure}
\usepackage{url}
\usepackage{hyperref}
\usepackage[labelfont=bf,labelsep=period]{caption}
\usepackage{lineno}
\usepackage{makecell}
\DeclareRobustCommand{\VAN}[3]{#2}
\let\VANthebibliography\thebibliography
\def\thebibliography{\DeclareRobustCommand{\VAN}[3]{##3}\VANthebibliography}

\def\arcsec{\hbox{$^{\prime\prime}$}}
\def\approxlt{\ifmmode \rlap{$<$}{}_{{}_{{}_{\textstyle\sim}}} \else%
$\rlap{$<$}{}_{{}_{{}_{\textstyle\sim}}}$\fi}

\def\xmm{XMM-{\it Newton}}
\def\swf{{\it Swift}}
\def\nic{{\it NICER}}

\def\arcsec{\hbox{$^{\prime\prime}$}}

\usepackage{hyperref}
\usepackage{academicons}
\usepackage{xcolor}
\usepackage{orcidlink}
\usepackage{tablefootnote}
\usepackage{longtable}

\title[AT2020ocn]{Tidal disruption event AT2020ocn: early--time X--ray flares caused by a possible disc alignment process}
\author[Cao et al.]{Z.~Cao,\orcidlink{0000-0002-0588-6555}$^{1,2}$\thanks{E-mail: z.cao@sron.nl},
P.G.~Jonker,\orcidlink{0000-0001-5679-0695}$^{2,1}$, D.R.~Pasham, \orcidlink{0000-0003-1386-7861}$^{3}$, S.~Wen,\orcidlink{0000-0002-0934-2686}$^{4,2}$, N.C.~Stone,\orcidlink{0000-0002-4337-9458}$^5$, A.I.~Zabludoff,\orcidlink{0000-0001-6047-8469}$^6$\\
$^1$SRON, Netherlands Institute for Space Research, Niels Bohrweg 4, 2333 CA Leiden, The Netherlands\\
$^2$Department of Astrophysics/IMAPP, Radboud University, P.O.~Box 9010, 6500 GL, Nijmegen, The Netherlands\\
$^3$Kavli Institute for Astrophysics and Space Research, Massachusetts Institute of Technology, Cambridge, MA 02139, USA\\
$^4$National Astronomical Observatories, Chinese Academy of Sciences, 20A Datun Road, Beijing 100101, China\\
$^5$Racah Institute of Physics, The Hebrew University, Jerusalem, 91904, Israel\\
$^6$Department of Astronomy and Steward Observatory, University of Arizona, 933 N. Cherry Ave., Tucson, AZ 85721, USA
}

\date{Accepted XXX. Received YYY; in original form ZZZ}

\pubyear{xxx}

\begin{document}
\label{firstpage}
\pagerange{\pageref{firstpage}--\pageref{lastpage}}
\maketitle

\begin{abstract}
A tidal disruption event (TDE) may occur when a star is torn apart by the tidal force of a black hole (BH). Eventually, an accretion disc is thought to form out of stellar debris falling back towards the BH. If the star's orbital angular momentum vector prior to disruption is not aligned with the BH spin angular momentum vector, the disc will be tilted with respect to the BH equatorial plane. The disc will eventually be drawn into the BH equatorial plane due to a combination of the Bardeen--Petterson effect and internal torques. Here, we analyse the X--ray and UV observations of the TDE AT2020ocn obtained by \swf{}, \xmm{}, and \nic{}. The X--ray light curve shows strong flares during the first $\approx100$~days, while, over the same period, the UV emission decays gradually. We find that the X--ray flares can be explained by a model that also explains the spectral evolution. This model includes a slim disc viewed under a variable inclination plus an inverse--Comptonisation component processing the slim disc emission. A scenario where the ongoing Lense--Thirring precession during the disc alignment process is responsible for the observed inclination variations is consistent with the data. In later observations, we find that the X--ray spectrum of AT2020ocn becomes harder, while the mass accretion rate remains at super--Eddington levels, suggesting the formation of a corona in line with accretion onto other compact objects. We constrain the BH mass to be $(7^{+13}_{-3})\times10^{5}$ ~M$_\odot$ at the 1$\sigma$ (68\%) confidence level.
\end{abstract}




\section{Introduction}
\label{sec:intro}

A star can be broken apart by tidal forces when approaching a black hole (BH), triggering a tidal disruption event (TDE; e.g., \citealt{hills1975possible,rees1988tidal}). A part of the stellar debris from the disrupted star will fall back towards the BH. The orbit of this fall-back material is expected to form an accretion disc \citep{rees1988tidal,evans1989tidal,ulmer1999flares}. Dozens of TDEs have been reported in the literature \citep{gezari2021tidal}, and the number of candidates is increasing rapidly, thanks to large--scale sky surveys such as \textit{Zwicky Transient Facility} (ZTF; \citealt{graham2019zwicky}), \textit{Asteroid Terrestrial-impact Last Alert System} (ATLAS; \citealt{tonry2018atlas}), and \textit{All Sky Automated Survey for Supernovae} (ASAS--SN; \citealt{shappee2014man}).

The disruption often leads to processes that generate optical/UV and X--ray emission \citep[e.g.,][]{bade1996detection,komossa2004huge,gezari2006ultraviolet,van2011optical,saxton2014x,van2020optical,saxton2020x}, which allows for the detection of massive BHs and the study of accretion processes. The thermal emission that is thought to originate in an accretion disc often dominates the TDE X--ray spectrum \citep[e.g.,][]{ulmer1999flares,lodato2011multiband,auchettl2017new}. In some cases, non--thermal powerlaw--like X--ray emission is also observed \citep[e.g.,][]{saxton2017xmmsl1,lin2017likely,wevers2019evidence,lin2020multiwavelength,jonker2020implications}. This non-thermal X-ray emission has been associated with the inverse--Comptonisation process where the thermal disc photons act as seed photons. While the late--time optical/UV emission (typically several hundred of days after the initial disruption) is consistent with originating from the disc \citep[e.g.,][]{van2019first,mummery2020spectral,wen2023optical}, the origin of the early--time optical and UV emission is still a matter of debate (e.g., see \citealt{roth2020radiative} for a review). One possibility is that UV photons are powered by the shocks (self--intersection shocks, or nozzle shocks) in the debris streams during the circularisation process, dissipating energy and angular momentum of the streams \citep[e.g.,][]{piran2015disk,shiokawa2015general,ryu2020measuring,andalman2022tidal,steinberg2022origins}. Another possibility is that the UV emission comes from a "reprocessing layer" that captures the X--rays emitted by the inner disc and re-emits their energy in the UV \citep[e.g.,][]{loeb1997optical,metzger2016wind,roth2018sets,dai2018unified,wevers2019evidence,bonnerot2020simulating}.

When the orbital angular momentum vector of the star prior to disruption is not aligned with the BH spin angular momentum vector, the disc plane might well be mis--aligned with respect to the BH equatorial plane \citep[e.g.,][]{stone2012observing,franchini2016lense}. Due to a combination of the Bardeen--Petterson effect \citep{stone2012observing} and internal torques \citep{franchini2016lense}, (the inner part of) this tilted disc will eventually be forced to align with the equatorial plane of a spinning BH. Predicted in theory and found in simulations, this disc alignment process manifests itself largely as Lense--Thirring precession, with the observed inclination angle of the disc varying during the process \citep[e.g.,][]{fragile2005hydrodynamic,franchini2016lense,zanazzi2019tidal,white2019tilted}. A varying disc inclination likely affects the broadband appearance of the source \citep[e.g.,][]{dai2018unified}. It has been proposed that the disc alignment is important in explaining the highly--variable jet features observed in jetted TDEs (e.g., Swift J164449.3+573451; \citealt{tchekhovskoy2014swift,liska2018formation}; see also \citealt{teboul2023unified}).

TDE studies can help test accretion theories in the super--Eddington regime. The mass accretion rate in the disc formed after the disruption can vary from highly super--Eddington to sub--Eddington levels \citep{strubbe2009optical,lodato2011multiband,guillochon2013hydrodynamical,metzger2016bright}. In the high/super--Eddington regime, energy advection across the BH horizon can no longer be neglected, and the disc geometry is different from the standard Shakura--Sunyaev geometrically thin disc, instead the disc is geometrically thick \citep{abramowicz1988slim}. In such cases, a "slim" disc model \citep[e.g.,][]{abramowicz1988slim,skadowski2009slim,skadowski2011relativistic} is more appropriate than the standard thin disc model \citep{shakura1973black}. Furthermore, TDEs are good laboratories for studying the spectral evolution associated with the transition from super-- to sub-- Eddington mass accretion rate. Many TDEs show spectral state transitions in the X--rays along their decay \citep[e.g.,][]{komossa2004huge,jonker2020implications,wevers2019evidence,cao2023rapidly}. Modelling the TDE X--ray spectrum allows to test whether such transitions appear under the same physical conditions as  spectral state transitions observed in other super--Eddington accretors (e.g., ultra--luminous X--ray sources [ULXs]; \citealt{gladstone2009ultraluminous,sutton2013ultraluminous,kaaret2017ultraluminous}).

The broadband source AT2020ocn (ZTF18aakelin) was first detected by ZTF in the optical on 2020-04-29 (Modified Julian Date, MJD~58968) and has been classified as a TDE candidate by \citet{gezari2020at2020ocn}. It is located at the centre of an otherwise quiescent, early type galaxy SDSS J135353.80+535949.7 at a redshift of $z$=0.0705. The $M-\sigma_*$ relation suggests a BH mass of $\sim10^{6.4\pm0.6}$ solar mass \citep{pasham2024lense}. Subsequent observations by the \textit{Neil Gehrels Swift} satellite revealed the source to be bright in the UV and the X--ray band \citep{gezari2020at2020ocn,miller2020recent}. \textit{Neutron star Interior Composition ExploreR} (\nic{}) started monitoring the source on 2020-07-11 (MJD~59041). Recently, \citet{pasham2024lense} discovered a $\sim17$-day quasi--periodicity modulating the X--ray flux of AT2020ocn as observed by \nic{} over the first $\sim130$~days. Various mechanisms can lead to this phenomenon, including a precessing accretion disc, as suggested by those authors. Therefore, spectral analysis with physical models is needed to see which mechanisms are consistent with the data.

In this paper, we use a standard $\Lambda$CDM cosmology with H$_{0}$ = 67.4 km~s$^{-1}$~Mpc$^{-1}$, $\Omega_{m}$ = 0.315 and $\Omega_{\Lambda}$ = 1 - $\Omega_{m}$ = 0.685 \citep{aghanim2020planck} when converting the redshift to the luminosity distance. Throughout the paper, $c$ is the speed of light, $G$ is the gravitational constant, $r$ is the radial coordinate measured from the BH centre, and $R_g$ is the gravitational radius $\frac{GM_{\bullet}}{c^2}$ for a BH of mass $M_{\bullet}$. We use $R_*$ and $M_*$, for the stellar radius and mass of the star prior to the disruption, respectively. $R_t$ is the tidal radius of the TDE, defined as $R_t=R_*(M_{\bullet}/M_*)^{1/3}$. 

Here we analyse the light curves in the UV and X--ray bands, and spectra of AT2020ocn obtained by \swf{}, \xmm{}, and \nic{}. The paper is structured as follows: In Section 2, we describe our data reduction method. In Section 3, we present the results from our analyses. In Section 4 we discuss the physical scenarios implied by our modelling. In Section 5, we present our conclusions.

\section{Data and data reduction}
\label{sc:data}

\subsection{\nic{}}
\label{sc:nicdata}

We started our \nic{} data analysis with the raw/level--1 files available on the HEASARC public archive\footnote{https://heasarc.gsfc.nasa.gov}. First, we reduced the data using the {\it nicerl2} task. Then, Good time intervals (GTIs) were produced with the default filters. We used the 3c50 background model \citep{Remillard2022} to extract background spectra on a per GTI basis. Following the recommendations by \citet{Remillard2022} we excluded GTIs that do not pass the level-3 filtering. For more details of our procedure please see: \cite{2023NatAs...7...88P,2018cowpasham}. Both the background and the source$+$background spectra of \nic{} are re--binned by the optimal--binning algorithm (\citealt{kaastra2016optimal}; ftool command \texttt{ftgrouppha}). Also, we ensure that both the background and the source$+$background spectra have a minimum of 1 count per bin (with parameter \texttt{grouptype} in \texttt{ftgrouppha} set to \texttt{optmin}).

Due to the super--soft X--ray nature of the source, \nic{} source counts are significantly below the background counts roughly above 1.1~keV for most of its observations. Therefore, we ignore \nic{} bands above 1.1~keV in this study. Meanwhile, to avoid spectra dominated by the noise in the background and uncertainties in estimating the background level, we exclude \nic{} observations where the source counts rate is lower than the background counts rate in 0.3-0.5~keV band. In this way, 1010 out of 1125 epochs of \nic{} observations remain.

When performing spectral analysis on \nic{} spectra, systematic errors of 1.5\%\footnote{identical to the systematic errors applied by the NICER data reduction task \texttt{niphasyserr}. See https://heasarc.gsfc.nasa.gov/docs/nicer/analysis\_threads/spectrum-systematic-error/} in the 0.3--1.1~keV band are added to the spectra using the "\texttt{systematic}" command in the \texttt{XSPEC} package (\citealt{arnaud1996xspec}; version 12.13.0c). We adjust the fitting energy range of each \nic{} spectrum using the \texttt{ignore} command in \texttt{XSPEC}, to discard the energy bins of the hard spectral tail where the source flux is lower than the background flux. The number of the discarded energy bins differs from epoch to epoch. Meanwhile, to have sufficient bins to fit a 2--parameter fit--function, we further require the spectrum to have a fitting energy range from 0.3~keV to at least 0.6~keV ($>$3 bins) to be considered for the fit procedure. In the analysis, data are considered to be consistent with the fit function if C-stat/d.o.f.~$<2$.

\subsection{XMM-{\it Newton}}
\label{sc:xmmdata}

\begin{table*}
\centering
\caption[]{\xmm{} observations of AT2020ocn analysed in this work. The exposure time is the time remaining after filtering for epochs of enhanced background count rates. The average count rates of the source$+$background spectra are given in the energy ranges 0.3--1.1~keV for XMM\#1 \& 2, and in 0.3--10.0~keV for XMM\#3. We also list in the last column the source counts calculated by subtracting the estimated number of background counts in the source extraction region.}
\begin{tabular}{ccccccc}
\hline
Satellite  & ObsID (Label)    & Date   & Exposure (ks) & Source region & Count rate (cts/s)  &  Est.~Source counts (cts) \\
\hline
XMM-Newton & 0863650101 (XMM\#1) & 2020-07-18 & 46 & annulus (15\arcsec{}-30\arcsec{}) & $(4.5\pm0.1)\times10^{-2}$&1517\\
           & 0863650201 (XMM\#2) & 2020-07-21 & 47 & annulus (15\arcsec{}-30\arcsec{}) & $(6.3\pm0.1)\times10^{-2}$&2833\\
           & 0872392901 (XMM\#3) & 2021-05-15 & 42 & circular (30\arcsec{}) & $(67.6\pm0.4)\times10^{-2}$&27546\\
\hline
\end{tabular}
\label{tb:xmmda}
\end{table*}

AT2020ocn was observed by \xmm{} on three occasions during the \nic{} monitoring. The observations are identified by their ID: 0863650101 (XMM\#1), 0863650201 (XMM\#2), and 0872392901 (XMM\#3). For the \xmm{} data reduction, we use HEASOFT (version 6.31.1) and SAS (version 20.0.0) with the calibration files renewed on October 25th, 2022 (CCF release: XMM-CCF-REL-391). During XMM\#2, the observation of two MOS detectors was interrupted for calibration purposes. Therefore, for consistency, we do not use the MOS data. We also do not use the RGS data, because the signal--to--noise ratio in the RGS detectors is too low. We use the SAS command {\sc epproc} to process the Science 0 data from \xmm/EPIC-pn. We employ the standard filtering criteria\footnote{https://www.cosmos.esa.int/web/xmm-newton/sas-thread-epic-filterbackground} for EPIC-pn data, where we require that the 10--12 keV detection rate of pattern 0 events is $<$ 0.4 counts s$^{-1}$. This way the data are cleared from periods with an enhanced background count rate. We use a circular source region of 30\arcsec{} radius centred on the source for the spectral counts extraction, corresponding to a $\sim$90\% energy fraction encirclement for a point source. Using the SAS command {\sc epatplot}, we check for the presence of photon pile--up, and find that XMM\#1 and XMM\#2 are suffered by pile--up while XMM\#3 does not. To clean the spectra from the pile--up effect, we use an annulus region of 15\arcsec{} inner radius and 30\arcsec{} outer radius for the source counts extraction of XMM\#1 and XMM\#2. We find no pile--up effect in the data when using such an annulus source region. In all three \xmm{} observations, the background spectral counts are extracted from apertures close to the source on the same EPIC-pn detector and free from other bright sources. We use a circular region of 50\arcsec{} radius for the background extraction. Using the \texttt{specgroup} command in SAS, we re--bin both the background and the source$+$background spectra of \xmm/EPIC-pn to have a minimum of 1 count per bin, while the oversampling factor is 3. We summarise the \xmm{} data used in this paper in Table~\ref{tb:xmmda}.

In this study we focus on the 0.3--10.0~keV band for the \xmm{}/EPIC-pn data. We find in XMM\#1 and XMM\#2 the background counts dominate the source$+$background spectrum $\gtrsim1$~keV. Therefore, we discard data above 1.1~keV during the analysis of XMM\#1 and XMM\#2, which is consistent with our treatment to the \nic{} data.  

\subsection{\swf{}}

We complement our study of the X-ray flares of AT2020ocn with the UV data from the UVOT instrument \citep{roming2005swift} on--board \swf{} satellite. We reduce the archived \swf{}/UVOT images of AT2020ocn obtained between 2020-06-25 (MJD~59025) and 2021-06-22 (MJD~59387) using the {\sc uvotproduct} task. We use a circular source region centred on the coordinates of AT2020ocn provided in the SIMBAD astronomy database\footnote{http://simbad.u-strasbg.fr/simbad/}, using a standard radius of 5 arcsec as suggested by the \swf{} team\footnote{https://www.swift.ac.uk/analysis/uvot/mag.php}. We use an annulus centred on the source as the background region, with an inner radius of 10 arcsec and an outer radius of 25 arcsec. We also extract the soft X-ray light curve from the \swf{}/XRT instrument using the online XRT tool\footnote{https://www.swift.ac.uk/user$\_$objects/} \citep{evans2009methods}, to compare with the X-ray behaviour of AT2020ocn as seen by \nic{}.

\vspace{10mm} 

Throughout this paper, we carry out the spectral analyses using the \texttt{XSPEC} package (\citealt{arnaud1996xspec}; version 12.13.0c). We use Poisson statistics (\citealt{cash1979parameter}; \texttt{C-STAT} in \texttt{XSPEC}). In this paper, we quote all the parameter errors at the 1$\sigma$ (68\%) confidence level, assuming $\Delta$C-stat=1.0 and $\Delta$C-stat=2.3 for single-- and two--parameter error estimates, respectively. In the figures the spectra are re-binned for plotting purposes only. In all fits we perform in this paper, we include the Galactic absorption using the model \texttt{TBabs} \citep{wilms2000absorption}, and we fix the column density $N_{H,G}$ to its measured value at the host direction of $1\times10^{20}{\rm cm}^{-2}$ \citep{schlafly2011measuring}. No intrinsic absorption is found by any of our fits in this paper. With the \texttt{energies} command in \texttt{XSPEC}, in all analysis we take a logarithmic energy array of 1000 steps from 0.1 to 1000.0~keV for model calculations instead of energy arrays from the response file, to be self--consistent and to correctly calculate the Comptonisation model when needed (see Section~\ref{sc:physicalmodels}). When needed, we use the Akaike information criteria (AIC; \citealt{akaike1974new}) to investigate the significance of adding model components to the fit--function, which is calculated by $\Delta$AIC$=-\Delta C +2\Delta k$ ($C$ is the C-stat and $k$ is the degree--of--freedom; \citealt{wen2018comparing}). The $\Delta$AIC$>$5 and $>$10 cases are considered a strong and very strong improvement, respectively, over the alternative model. 

For each spectrum of \xmm{}/EPIC-pn or \nic{}, we first fit the background spectrum with a phenomenological model. When fitting the source$+$background spectrum, we then add the best--fit background model to the fit function, fixing the parameters of the background model to their best--fit values determined from the fit to the background--only spectrum. The best--fit background model for \xmm{}/EPIC-pn data varies from epoch to epoch, consisting of 1 Gaussian components and 2 power--law components for XMM\#1 and XMM\#2, or 3 power--law and 1 Gaussian components for XMM\#3 that detects photons of higher energies, accounting for the background continuum and background fluorescence lines \citep[e.g.][]{katayama2004properties}. Meanwhile, we find the background spectra of \nic{} data can be described by a model consisting of 2 Gaussian components and 2 power--law components. The full--width half--maximum or FWHM $\sigma_{\rm gauss}$ is set to 0.001~keV for all the Gaussian components with a FWHM lower than the spectral resolution of \xmm{}/EPIC-pn or \nic{}. In the following, when studying the source$+$background data, we refer only to the part of the fit function that describes the source as \textit{fit function}.

\section{Results}

\subsection{Long--term light curve of AT2020ocn}

\begin{figure*}
    \centering
    \includegraphics[width=0.8\textwidth]{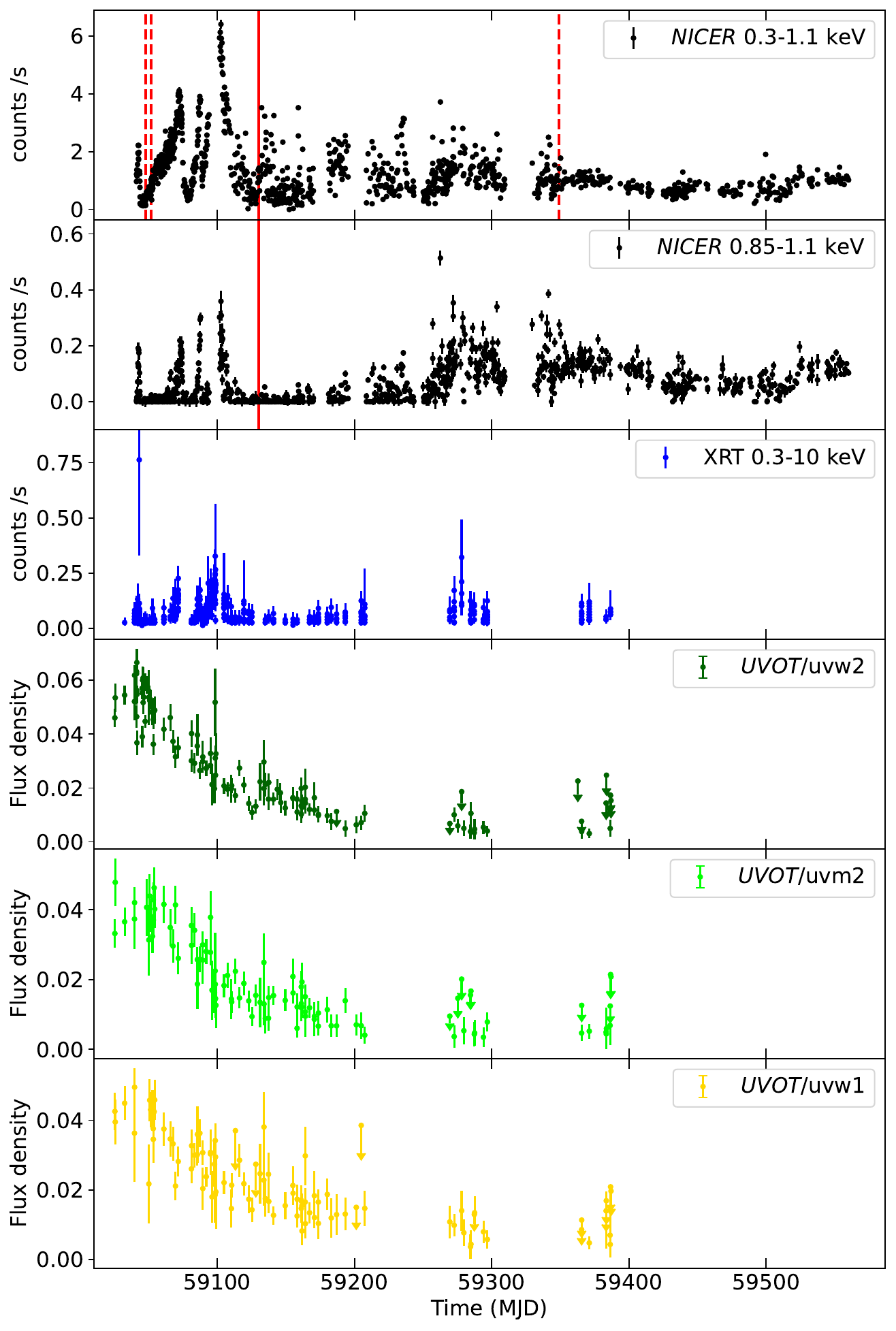}
    \caption{Long--term light curve of AT2020ocn as observed by \nic{} and \swf{}. From top to bottom: \nic{} count rate in the 0.3-1.1 keV band; \nic{} count rate in the 0.85-1.1 keV band; \swf{}/XRT count rate in the 0.3-10 keV band; the flux densities are in units of $10^{-26}$~erg/s/cm$^2$/Hz for the \swf{}/UVOT light curves in the uvw1, uvm2, and uvw2 filters. The x--axis denotes Modified Julian Date (MJD). Dashed lines in the top panel mark the times of the \xmm{} observations (in this paper we refer to these as XMM\#1, XMM\#2, and XMM\#3 in chronological order). The solid line marks MJD~59130, the date we use in this paper to separate the so--called early and late periods.}
    \label{fig:lc}
\end{figure*}

We show the long--term UV and X--ray light curves of AT2020ocn in Fig.~\ref{fig:lc}. Based on the \nic{} data, we find that the behaviour of the X--ray emission of AT2020ocn can be divided into two stages: an early period when strong X-ray flares are present (Modified Julian Date [MJD] $\lesssim$59130) and a late period of more gradual changes (MJD$\gtrsim$59130). There are 375 epochs for the early period, and 635 for the late period. The \nic{} hard X--ray (0.85-1.1~keV) count rate is low outside the flares in the early epoch data. The flares cannot be explained by background fluctuations (Fig.~\ref{fig:nicbkg}). In the late period, the light curve in the hard band shows a re--brightening around MJD~59300, with strong variations in the count rate. The X-ray light curve does not show a gradual decay in general. The \swf{}/XRT also detected the early X--ray flares. Meanwhile, the UV flux of AT2020ocn in all of the three UV bands of \swf{}/UVOT (uvw1, uvm2, and uvw2 band) decreases gradually with time, showing no evidence for flares such as those in the X-ray light curve. In other words, the X--ray and the UV light curves seem to be decoupled in the case of AT2020ocn.

\subsection{Long--term spectral evolution of AT2020ocn}

\begin{figure*}
    \centering
    \includegraphics[width=0.8\textwidth]{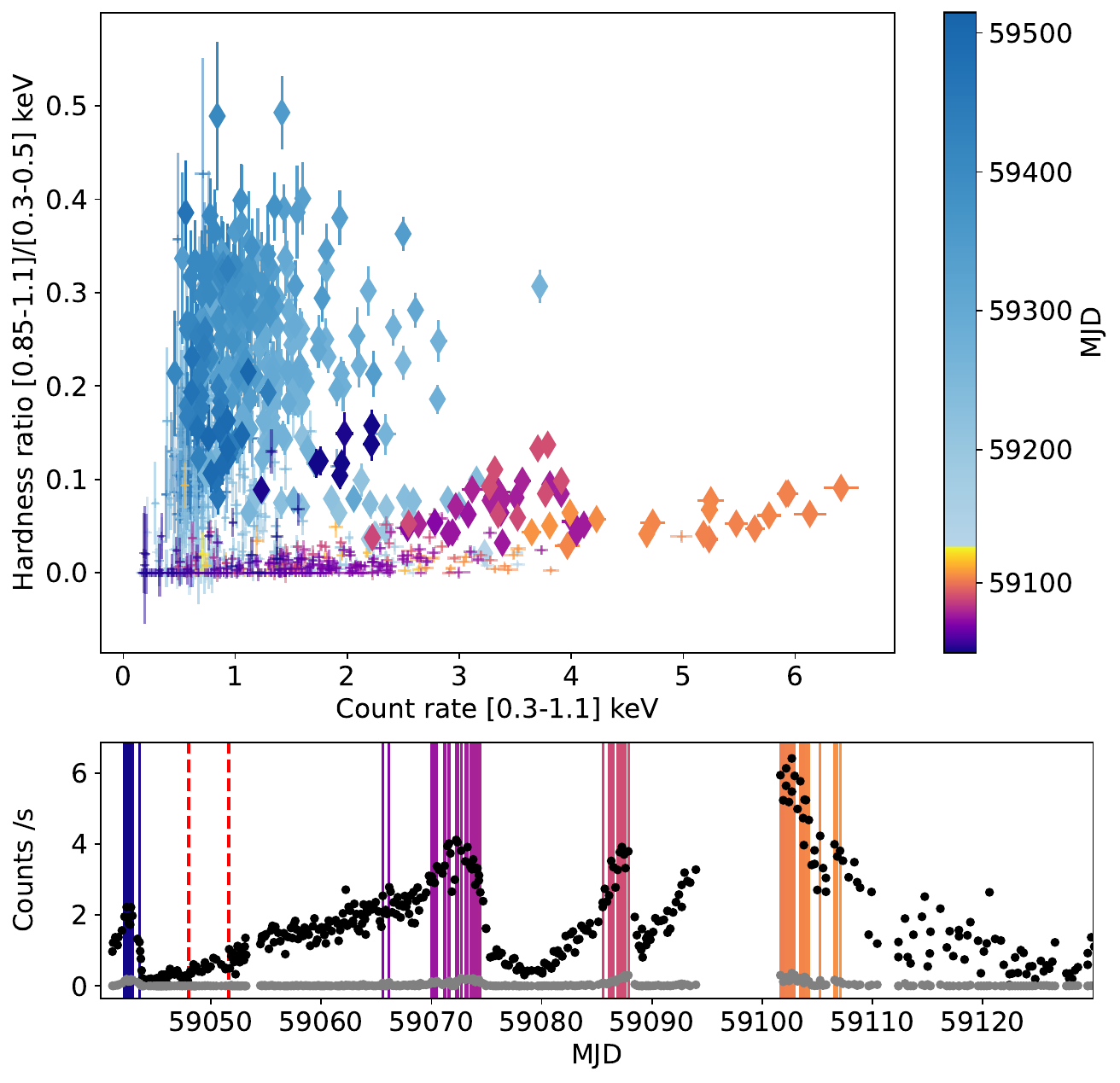}
    \caption{\textit{Upper}: We show here the \nic{} hardness ratio vs.~the \nic{} count rate in the 0.3-1.1~keV band. We define the spectral hardness ratio as the ratio between the count rate in the 0.85-1.1~keV band and that in the 0.3-0.5~keV band. The shown data are colour--coded by their observation time in MJD. We use different colours to show the hardness evolution during the early period (MJD$<$59130; from purple to yellow), and during the late period (MJD$>$59130; from light blue to dark blue). Epochs highlighted with diamond markers have the source count rate higher than the background count rate in the 0.85-1.1~keV band, so that their hardness ratios determined are least affected by the uncertainties in the \nic{} background estimation. \textit{Bottom}: Zoom-in of the \nic{} early--time light curve (i.e.,~data obtained before 59130~MJD) also shown in the {\it top} panel in Fig~\ref{fig:lc}. The black and grey data show the source count rates in the 0.3-1.1~keV and 0.85-1.1~keV band, respectively. The solid coloured vertical lines mark the observation times of the corresponding data highlighted by diamond markers in the upper panel, following the same colour scheme. From left to right, the red dashed lines mark the observation time of XMM\#1 and XMM\#2, respectively.}
    \label{fig:hid}
\end{figure*}

We study the spectral evolution of the source using the spectral hardness ratio in the \nic{} bands. We define the spectral hardness ratio as the ratio between the count rates in the 0.85-1.1~keV and the 0.3-0.5~keV bands. In analogy with the hardness--intensity diagram used often in X--ray binary studies, we present the hardness ratios of the \nic{} data of AT2020ocn as a function of the broadband count rate (0.3-1.1~keV) in Fig.~\ref{fig:hid}. We find that the evolution of the hardness ratio is different for the early and the late period. Generally, during the early period the source spectrum is softer and the observed flux is higher. The hardness evolution during the 4 flares traces out a different region of the hardness-intensity diagram, showing a harder--when--brighter pattern within each flare. During the late period, the source hardness ratio is higher while the observed flux is lower. 

\begin{figure}
    \centering
    \includegraphics[width=\linewidth]{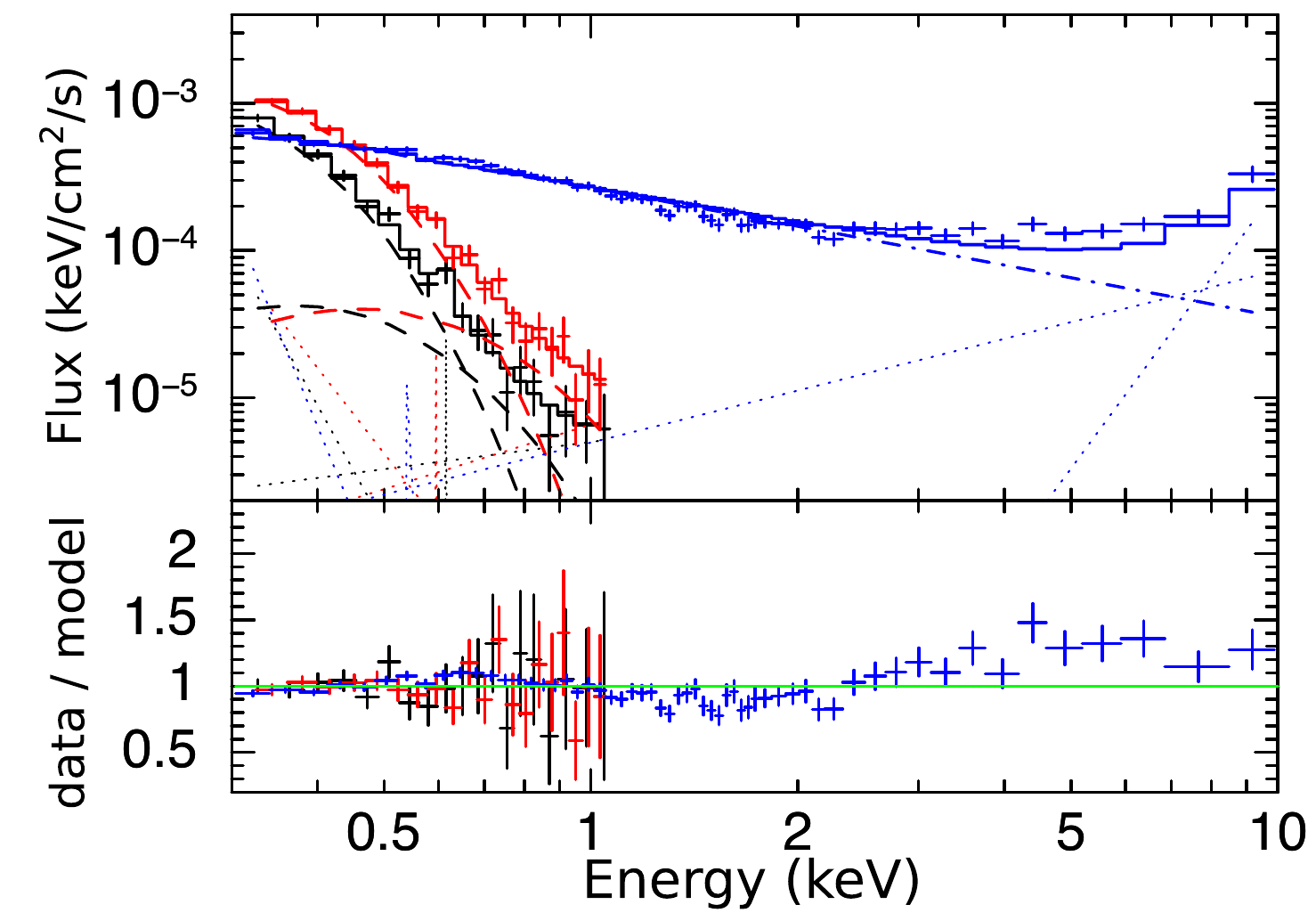}
    \caption{Top panel: We show the EPIC-pn source$+$background spectra of XMM\#1 (\textit{black}), XMM\#2 (\textit{red}), and XMM\#3 (\textit{blue}). The solid, dashed, dot--dashed and dotted lines represent the best--fit total models using phenomenological source models, the black body models used for XMM\#1 \& 2, the power--law model used for XMM\#3, and the contribution from the background as determined from fitting extracted spectra from background--only data separately, respectively. The best--fit background power--law indices and Gaussian parameters have been held constant during the fit to the source$+$background spectra. 
    Bottom panel: The ratio between the observed number of counts in each spectral bin (data; black, red and blue points in the top panel) and the best--fit predicted number of counts in each spectral bin (model; solid lines in the top panel) is shown.}
    \label{fig:xmmbb+po}
\end{figure}

We here use phenomenological model fits to the \xmm{}/EPIC-pn spectra to investigate the spectral changes between the early and late spectra in some detail. Of the three \xmm{} observations obtained during the \nic{} monitoring period (marked by dashed lines in Fig.~\ref{fig:lc}), the observation XMM\#1 and XMM\#2 are in the early period while XMM\#3 falls in the late period. We find that the XMM\#1 and the XMM\#2 spectra can be fitted well together (C-stat/d.o.f.~$=20.5/31$) with a fit--function comprised of two black body models (Fig.~\ref{fig:xmmbb+po}, we used the black body model \texttt{zbbody} in XSPEC syntax). However, XMM\#3 cannot be fitted well with such a fit function. Instead, it can be fitted with a power law with a photon index $\Gamma=2.89\pm0.01$ (C-stat/d.o.f.~$=316.5/166$). Fig.~\ref{fig:xmmbb+po} shows the different \xmm{}/EPIC-pn spectra. The XMM\#3 spectrum is much harder than the other two spectra. This spectral difference between the early and late spectra observed by \xmm{} is consistent with our findings based on the \nic{} data (Fig.~\ref{fig:hid}).

\subsection{Spectral analysis of X-ray data}
\label{sc:physicalmodels}

\subsubsection{XMM\#1 and XMM\#2}
\label{xmm12}

\begin{figure}
    \centering
    \includegraphics[width=\linewidth]{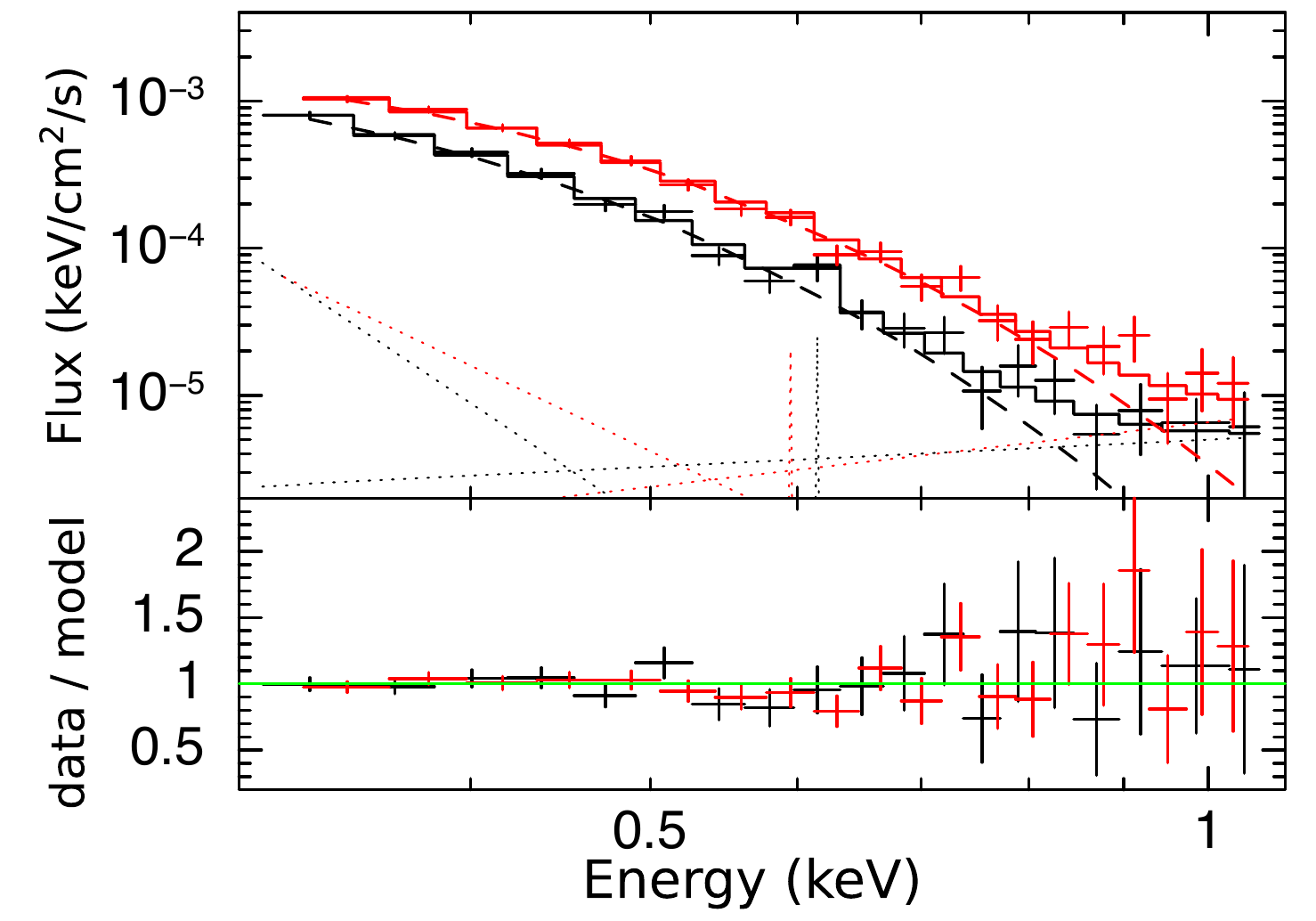}
    \caption{Top panel: the EPIC-pn XMM\#1 (\textit{black}) and XMM\#2 (\textit{red}) spectra of AT2020ocn fitted by a fit--function comprised of the following model components: \texttt{TBabs*slimdz}. The solid, dashed, and dotted lines represent the best--fit total model, the slim disc emission, and the contribution from the background as determined from fitting extracted spectra from background-only data separately, respectively. The best-fit background power--law indices and Gaussian parameters have been held constant during the fit to the source$+$background spectra. Bottom panel: We show the ratio between the observed number of counts (data; red and black points in the top panel) and the best-fit predicted number of counts in each spectral bin (model; red and black solid lines in the top panel).}
    \label{fig:xmmslim}
\end{figure}

\begin{figure}
    \centering
    \includegraphics[width=\linewidth]{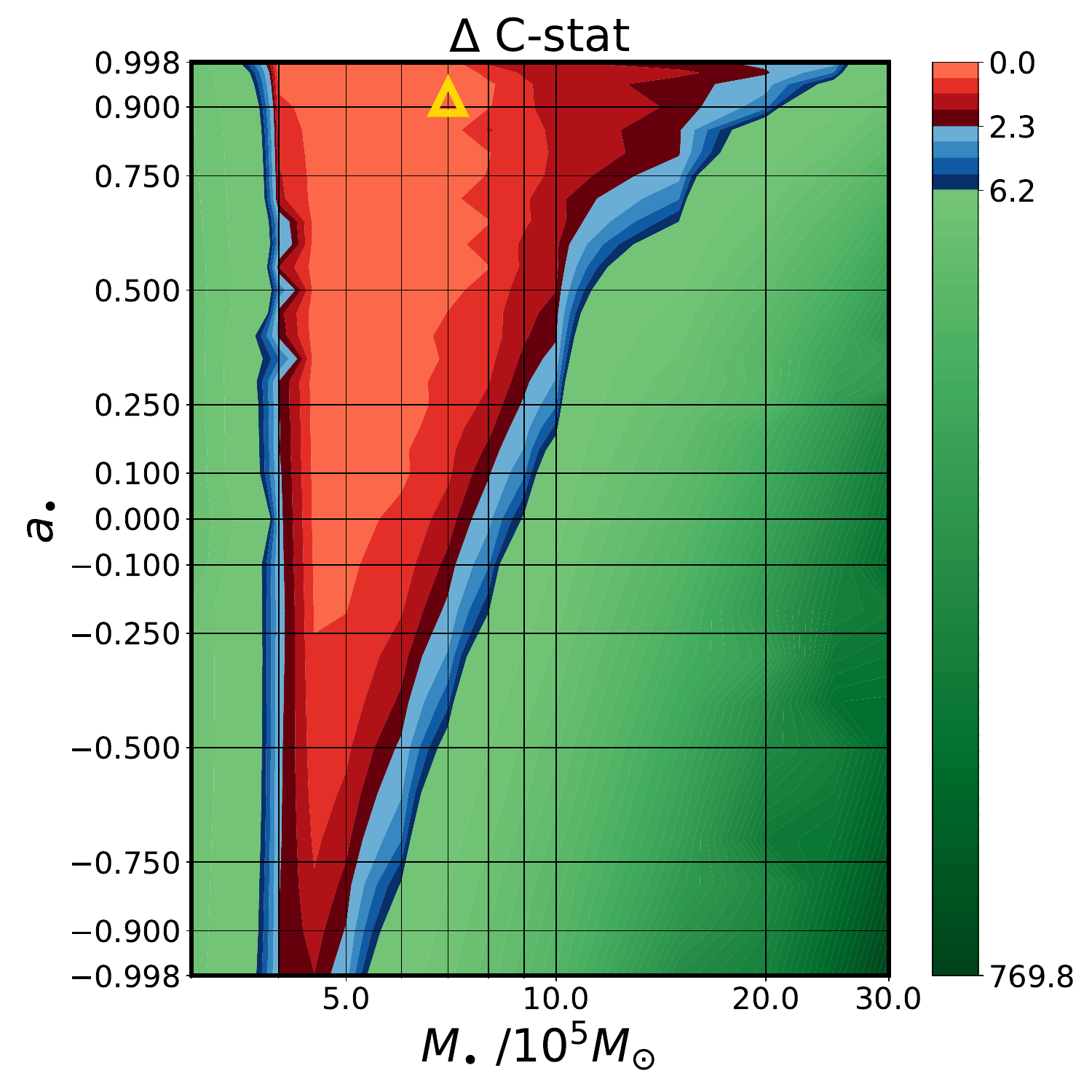}
    \caption{Constraints on $M_{\bullet}$ and $a_{\bullet}$ from the slim disc model-fit to the joint XMM\#1 \& 2 spectra. We calculate the $\Delta$C-stat across the $\{M_{\bullet}$, $a_{\bullet}\}$ plane. The best--fit point with the lowest C-stat is marked by a yellow triangle. Areas within 1$\sigma$ and 2$\sigma$ for two--parameter error estimations are filled by red and blue colours, respectively. At 1$\sigma$ for the two-parameter fits, $M_{\bullet}$ is constrained to be ($7^{+13}_{-3})\times10^{5}$~M$_\odot$.}
    \label{fig:ctxmm}
\end{figure}

To constrain parameters (such as the BH mass and spin, and the accretion rate) of AT2020ocn, we use the slim disc model \texttt{slimdz} \citep[][]{wen2022library} to simultaneously fit the two XMM\#1 and XMM\#2 spectra, allowing the mass accretion rate $\dot m$ to vary between epochs. We find a good fit with the slim disc model (C-stat/d.o.f.~$=26.4/34$; \texttt{XSPEC}'s syntax \texttt{"TBabs*slimdz"}). As stated in Section~\ref{sc:data}, we fix the column density of the Galactic absorption ($N_{H,G}$ in \texttt{TBabs}) to to its measured value in the host's direction $1\times10^{20}{\rm cm}^{-2}$ \citep{schlafly2011measuring}. We will keep using this $N_{H,G}$ value and fix it during the fits for all the following fit--functions considered in this paper. For a 1$\sigma$, single--parameter error estimate, the best--fit value for the BH mass $M_{\bullet}$ is $(7\pm1)\times10^{5}$~$M_{\odot}$, and for the inclination it is $74^{+1}_{-11}$ degrees. The BH spin $a_{\bullet}$ is constrained to have a lower limit of 0.25. We present the best--fit slim disc model in Fig.~\ref{fig:xmmslim}, and the parameter constraints in Table~\ref{tb:1and2par}. The accretion rate $\dot m$ at these two epochs do not differ from each other by more than 1$\sigma$ error range. By investigating the $\Delta$C-stat across the $\{M_{\bullet}$, $a_{\bullet}\}$ plane, we find a degeneracy between the BH mass and BH spin (Fig.~\ref{fig:ctxmm}): the lower limit on $a_{\bullet}$ increases with increasing $M_{\bullet}$. In general, $a_{\bullet}$ cannot be constrained. Specifically, the 1$\sigma$ lower limit on $a_{\bullet}$ for a two--parameter error estimate, at the best--fit BH mass value, is $a_{\bullet}>-0.1$. We also find an equivalently--good fit with not $\dot m$ but $\theta$ varying between the two epochs (C-stat/d.o.f.~$=26.1/34$; Table~\ref{tb:1and2par}). In this test case, the BH mass is $M_{\bullet}=(7\pm1)\times10^{5}$~$M_{\odot}$ and $a_{\bullet}>0.41$; the inclination $\theta$ at these two epochs is consistent with being constant within 1$\sigma$ ($74^{+1}_{-9}$ degree for XMM\#1, and $72^{+1}_{-9}$ degree for XMM\#2). 

\subsubsection{NICER spectra from the early flaring period}
\label{sc:nicscan}

\begin{figure*}
    \centering
    \includegraphics[width=\linewidth]{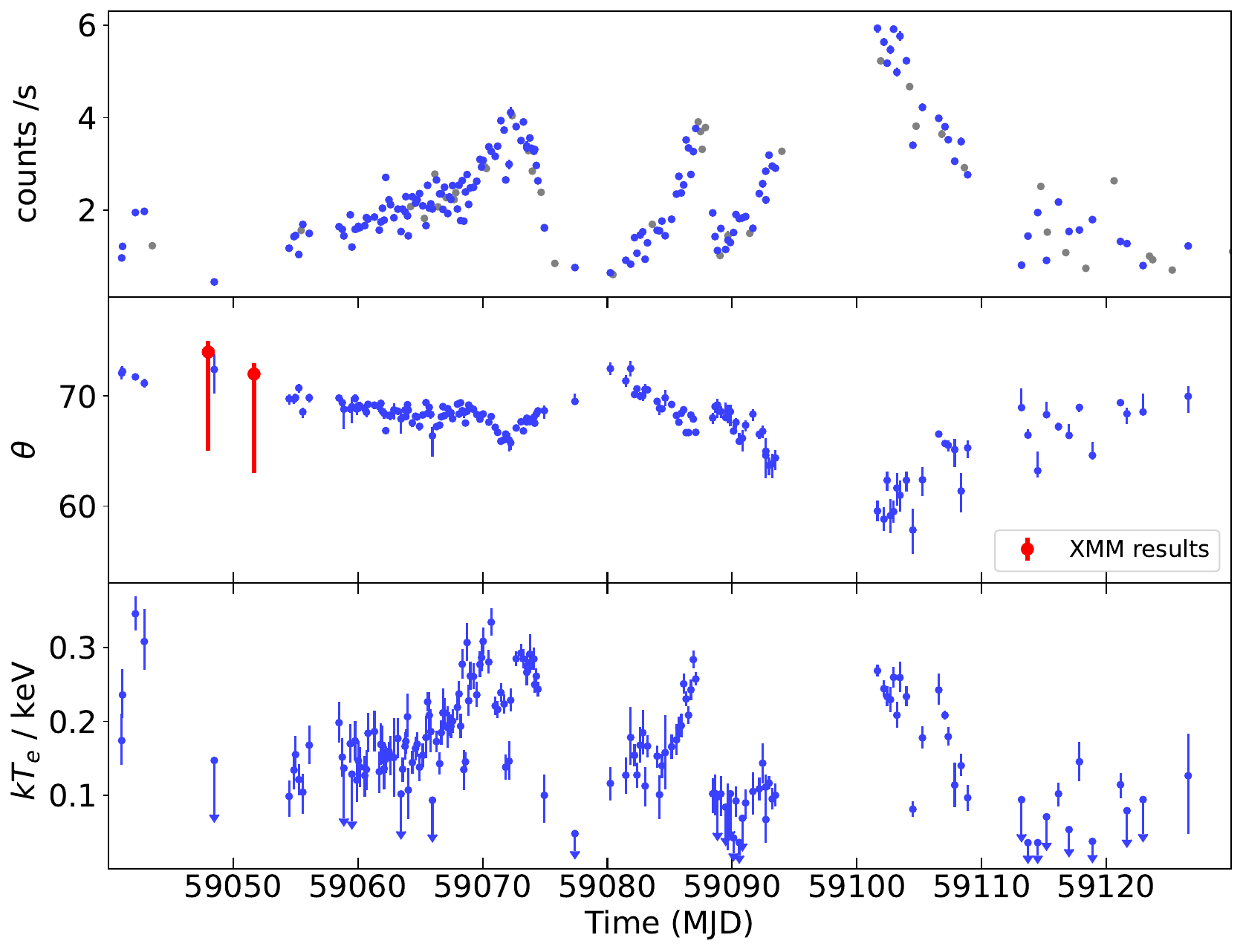}
    \caption{\textit{Top panel}: The \nic{} 0.3-1.1~keV early--time source light curve. We only consider the epochs where the source flux stays above the background level from 0.3~keV to at least 0.6~keV (a total of 206 epochs in the early period). Spectra at epochs marked by blue dots are well--fit (C-stat/d.o.f.~$<2$) by the model in this analysis, while grey dots mark spectra that have C-stat/d.o.f.~$>2$. \textit{Middle panel}: Inclination $\theta$ constraints derived from the \nic{} spectra obtained before MJD~59130. The fit--function is comprised of the following model components: \texttt{TBabs*(thcomp*slimdz)}. We fit each of the 206 spectra individually allowing the inclination $\theta$, and the temperature of the Comptonising medium $kT_e$ to vary. We fix $M_{\bullet}=7\times10^{5}$~$M_{\odot}$ and $a_{\bullet}=0.9$ based on the results of the spectral fits to the \xmm{} data (Fig.~\ref{fig:ctxmm}). The slim disc accretion rate $\dot m$ and the optical depth of the Comptonising medium are held fixed at a value of 30~$\dot m_{\rm Edd}$ and $\tau=20$, respectively. We show the fit parameter values for the 165 out of the 206 spectra where the C-stat/d.o.f.~$<2$ (the results from the blue points in the top panel). The inclinations constrained from the joint--fit to the XMM\#1 and XMM\#2 X--ray spectra are marked with the red dots.
    \textit{Bottom panel}: Constraints on the electron temperature $kT_e$ derived from the same fitting procedure described above.}
    \label{fig:nicconstraints}
\end{figure*}

Given the results of the spectral fits to the XMM\#1 and XMM\#2 data, we try to fit the individual \nic{} spectra of the early flaring period using the same fit--function (\texttt{TBabs*slimdz}). We aim to test whether allowing the values of one or more parameters of the slim accretion disc to vary can explain the spectral and flux variability observed by \nic{} over the early period. According to our selection criteria of \nic{} data as described in Section~\ref{sc:nicdata}, a total of 206 epochs in the early period are considered for the spectral analysis. We list these epochs in Table~\ref{tb:niceobs}.  

For the slim disc fit function, we fix the BH mass $M_{\bullet}$ to the best--fit value obtained from the joint XMM\#1 and XMM\#2 spectral fit using the same fit--function ($7\times10^{5}$~$M_{\odot}$), and we fix the BH spin $a_{\bullet}=0.9$. This spin value corresponds to the best--fit slim disc model to the joint XMM\#1 \& 2 spectra given our choice of the BH mass value (Fig.~\ref{fig:ctxmm}).

We first test whether the spectral evolution of the early period \nic{} data can be explained by a slim disc varying its accretion rate $\dot m$ as well as its inclination angle $\theta$ with respect to our line-of-sight. We find that for only 108 out of the 206 \nic{} early spectra this procedure gives a C-stat/d.o.f.~$<2$ (Fig.~\ref{fig:nicslim}). The slim disc model fails to fit most of the spectra at the peak of each flare. Specifically, from the fit residuals we find the spectral hardening at the flare peaks makes the spectrum deviate from a slim disc model (e.g., Fig.~\ref{fig:nicexamp}). We conclude that the X-ray flares of AT2020ocn in the early period can not be fully explained by only varying the disc accretion rate $\dot m$ and the inclination angle $\theta$. 

A hard spectral component additional to a disc continuum in the X--ray spectra of BH accretion systems has been interpreted before to be due to inverse--Comptonisation (IC; e.g., \citealt{belloni2010states,kubota2019modelling,mummery2021hard}), where high--energy electrons scatter the soft disc photons to higher energies. This IC component is commonly used to explain the harder--than--disc BH spectra in many TDEs or ultra--luminous X--ray sources that have been proposed to be at near--/super-- Eddington accretion rate \citep[e.g.,][]{magdziarz1998spectral,gladstone2009ultraluminous,saxton2019xmmsl2,wevers2021rapid,yao2022tidal}. Therefore, we test if the spectra at the peak of the flares can be fit well when including an IC component with seed photons coming from the slim disc, using the model \texttt{thcomp} (\citealt{zdziarski2020spectral}; the \texttt{thcomp} model parameterises the up--scattered spectra through the Thomson optical depth $\tau$ and the electron temperature $kT_e$. The total fit function in \texttt{XSPEC}'s syntax is \texttt{"TBabs*thcomp*slimdz"}). We fix the covering fraction of \texttt{thcomp} to unity, so that all seed photons go through the Comptonising medium. Given the data quality we find parameter degeneracies in most epochs between the IC electron temperature $kT_e$ and the IC optical depth $\tau$. The best--fit $\tau$ is typically $\gtrsim10$. We therefore fix $\tau=20$ during the fit so that the IC component is parameterised by a single free parameter ($kT_e$). 

We find that for 165 out of the 206 \nic{} early spectra the fit procedure gives a C-stat/d.o.f.~$<2$ by assuming a similar $\dot m$ as determined in XMM\#1 \& 2 ($\dot m=30$~$\dot m_{\rm Edd}$). Each spectrum has at least 2~d.o.f. left to be fitted with a 2--parameter ($\theta$ and $kT_e$) fit function, and most of the not--well--fit spectra also give a C-stat/d.o.f.~close to 2 (Fig.~\ref{fig:stathist}). We present the evolution of the parameter constraints produced from this fit procedure in Fig.~\ref{fig:nicconstraints}. This slim disc$+$IC model can fit most of the spectra at flare peaks (e.g., Fig~\ref{fig:nicesoftexamp}). We find that the Comptonising medium becomes hotter ($kT_e\sim0.3$~keV) during the flares, while the $kT_e$ is lower ($kT_e\lesssim0.3$~keV) outside the flares. Meanwhile, no intrinsic absorption is needed to model any flares. 

Next, we investigated if variations in $\dot m$ and $kT_e$ can explain the early X--ray flares instead of variations in $\theta$ and $kT_e$. When fitting individual spectra, we fix $\theta$ to the best--fit value determined from XMM\#1 \& 2 ($74^{\circ}$) and let $\dot m$ and $kT_e$ free to vary. This fit procedure results in a much lower number of spectra to be well-fit by the slim disc$+$IC model (59 spectra fitted instead of 165 for the procedure where $\theta$ and $kT_e$ are free to vary). Letting also $\tau$ free to vary does not improve the fits. We conclude that the variation in inclination are likely to be important during the early period, while the $\dot m$ variation is not the leading factor causing the flares.

\begin{table*}
\renewcommand{\arraystretch}{1.5}
\centering
\caption{Parameter constraints and the fit statistics from our joint--fits to the XMM\#1 and XMM\#2 spectra. Parameter values held fixed during the fit are given inside square brackets. In this table we quote the parameter errors derived using $\Delta$C-stat=1.0 for single--parameter error estimates. The first joint--fit assumes a difference in the mass accretion rate $\dot m$ between the two epochs, while the second joint--fit assumes a difference in the inclination $\theta$.}
\begin{tabular}{c|c|c|cccc}
    \hline
    Fit--function & Epoch & \texttt{Tbabs} & \multicolumn{4}{c}{\texttt{slimdz}} \\
     & & $N_H/10^{20}$~cm$^{-2}$ & $\dot m/\dot m_{\rm Edd}$ & $\theta/^{\circ}$ & $M_{\bullet}/M_{\odot}$ & $a_{\bullet}$ \\
     \hline
    \texttt{Tbabs*slimdz} & XMM\#1 & [1.0] &$>17$ & $74^{+1}_{-11}$ & $(7\pm1)\times10^{5}$ &  $>0.25$ \\
     & XMM\#2 & =XMM\#1 & $19^{+173}_{-6}$ & =XMM\#1 & \multicolumn{2}{c}{=XMM\#1} \\
     \hline
     \multicolumn{6}{c}{C-stat/d.o.f.~$=26.4/34$}\\
     \hline
    \hline
    Fit--function & Epoch & \texttt{Tbabs} & \multicolumn{4}{c}{\texttt{slimdz}} \\
     & & $N_H/10^{20}$~cm$^{-2}$ & $\dot m/\dot m_{\rm Edd}$ & $\theta/^{\circ}$ & $M_{\bullet}/M_{\odot}$ & $a_{\bullet}$ \\
     \hline
    \texttt{Tbabs*slimdz} & XMM\#1 & [1.0] &$27^{+171}_{-4}$ & $74^{+1}_{-9}$ & $(7\pm1)\times10^{5}$ &  $>0.41$ \\
     & XMM\#2 & =XMM\#1 & =XMM\#1 & $72^{+1}_{-9}$ & \multicolumn{2}{c}{=XMM\#1} \\
     \hline
     \multicolumn{6}{c}{C-stat/d.o.f.~$=26.1/34$}\\
     \hline
\end{tabular}
\label{tb:1and2par}
\end{table*}

\subsubsection{The late--time spectrum observed in XMM\#3}

\begin{figure}
    \centering
    \includegraphics[width=\linewidth]{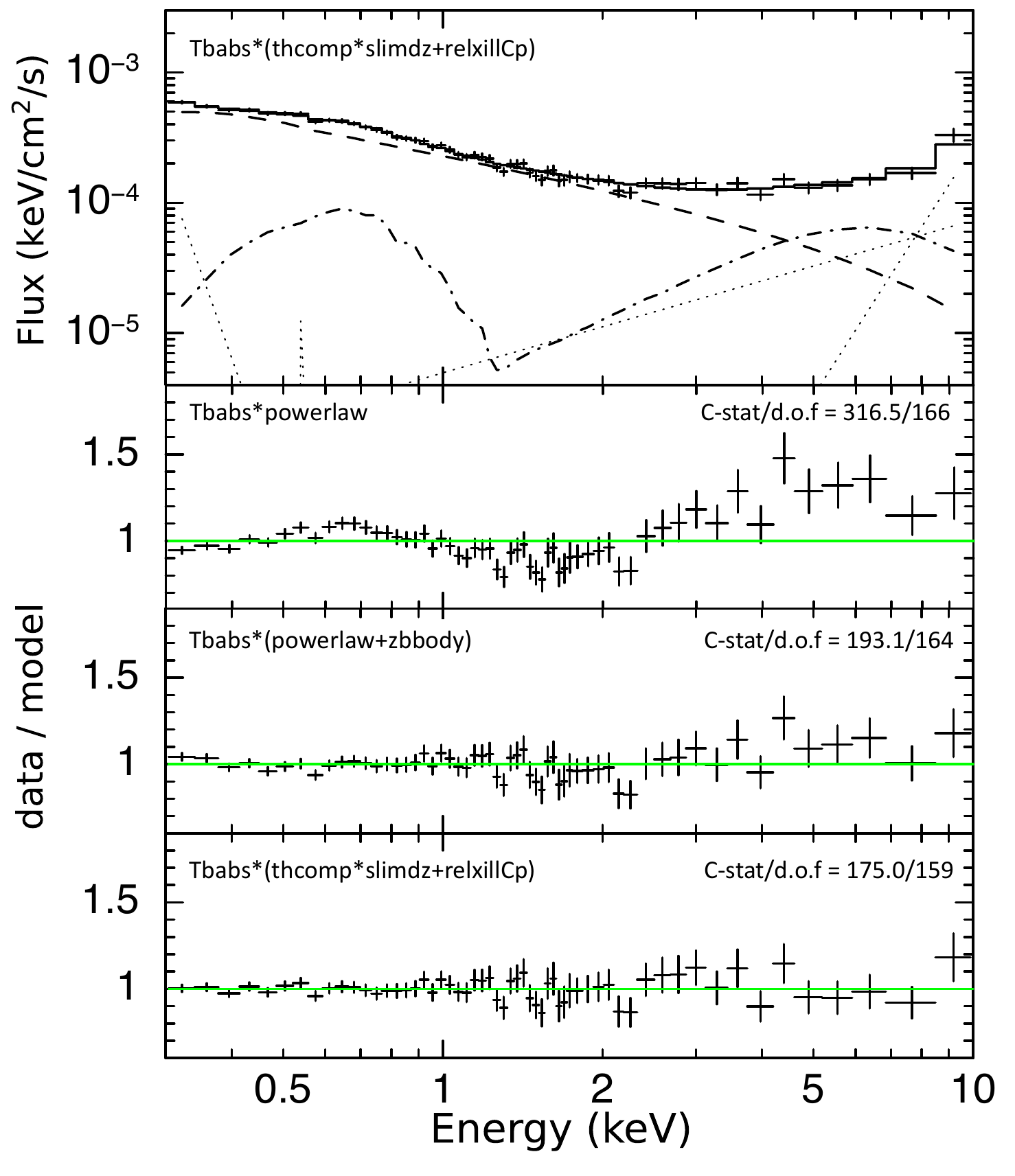}
    \caption{\textit{Top panel}: The X--ray spectrum and the best--fit \texttt{"TBabs*(thcomp*slimdz+relxillCp)} model for the XMM\#3 observation. The solid, dashed, and dot--dashed lines represent the best--fit total model, the coronal component \texttt{thcomp*slimdz}, and the corona reflection \texttt{relxillCp}, respectively. The dotted black lines show the contribution from the background as determined from fitting separately background--only spectra. The best--fit background power--law indices have been held constant during the fits to the source spectrum. \textit{Upper--middle panel}: The ratio between the observed number of counts (data) and the best-fit predicted number of counts in each spectral bin (model), for a source fit--function of \texttt{"TBabs*powerlaw"} to describe the XMM\#3 spectrum. \textit{Lower--middle panel}: The ratio between the data and the model for a source fit--function of \texttt{"TBabs*(powerlaw+zbbody)"} to describe the XMM\#3 spectrum.
    \textit{Bottom panel}: The ratio between the data and the model for the source fit--function of \texttt{"TBabs*(thcomp*slimdz+relxillCp)"} shown in the top panel. }
    \label{fig:xmm3}
\end{figure}

During the late--time period (MJD$>$59130) XMM-{\it Newton} observed AT2020ocn once (XMM\#3). The spectrum of XMM\#3 can be well-fit by a power law with index $\Gamma=2.89\pm0.01$ (\texttt{TBabs*powerlaw}; C-stat/d.o.f.~$=316.5/166$). However, there are trends in the residuals for this fit showing the model under--predicting the data systematically above 2~keV (Fig.~\ref{fig:xmm3}). The fit can be then significantly improved using a fit--function consisting of a power law and a black body (C-stat/d.o.f.~$=193.1/164$ and $\Delta$AIC~$=119.4$; \texttt{XSPEC}'s syntax \texttt{"TBabs*(powerlaw+zbbody)"}). The black body model with a temperature of $\sim$0.12~keV accounts for part of the continuum at the soft end ($<$2.0~keV; Fig.~\ref{fig:xmm3phe}). 

Besides these phenomenological models, we use a fit--function to fit the spectrum of XMM\#3 that contains \texttt{TBabs*thcomp*slimdz}. Like the case using a fit--function of only a power--law, this fit--function does not describe the XMM\#3 data well and has similar residuals. Part of the coronal emission can be reflected off the accretion disk, this reflection spectrum is calculated using the \texttt{relxillCp} model \citep[][]{dauser2014role,garcia2014improved}. In \texttt{relxillCp}, the disc is assumed to be a standard Shakura--Sunyaev thin disc \citep[][]{shakura1973black}, and the incident coronal emission is modelled by \texttt{nthcomp} (the now depreciated, stand--alone version of \texttt{thcomp}; \citealt{zdziarski1996broad,zdziarski2020spectral}), that assumes a multi--temperature black body seed spectrum. There are currently no reflection models using a slim disc for the disc seed photons, and so we use \texttt{relxillCp} to approximate the reflected emission off a slim disc. For this reason, one should be cautious when interpreting the results. 

Overall, the total fit--function in \texttt{XSPEC}'s syntax is \texttt{"TBabs*(thcomp*slimdz+relxillCp)"}. Given the mentioned inconsistency of disc assumptions between the two model components, we do not try to measure the $M_{\bullet}$ and $a_{\bullet}$ values by analysing XMM\#3 using this fit-function. Instead, we fix the $M_{\bullet}$ and $a_{\bullet}$ to their best--fit values from the analysis of XMM\#1 \& 2 (Fig.~\ref{fig:ctxmm}). The inclination $\theta$ shared between the \texttt{slimdz} and \texttt{relxillCp} models is free to float in the fit but it is required to be the same between models. The $kT_e$ shared between the \texttt{thcomp} and \texttt{relxillCp} models is also free to float in the fit but likewise it is required to be the same between models. We fix the Refl$_{\rm frac}$ parameter in \texttt{relxillCp} to be $-1$ so that the model only accounts for the reflected emission. Other parameters for \texttt{relxillCp} that we have held fixed are: the iron abundance $A_{\rm Fe}=1$ (in units of the solar abundance), the redshift $z=0.0705$, the disc inner radius which we assumed to be at the innermost--stable--circular--orbit (ISCO; $R_{\rm in}=-1$ in \texttt{relxillCp}'s syntax), the disc's particle density $\rho=10^{17}$~cm$^{-3}$ from order--of--magnitude estimations \citep{2015tdss.book.....S}, and the disc outer radius $R_{\rm out}=100R_g$ since that is the typical scale of a TDE disc for a BH of $1\times10^{6}$~$M_{\odot}$ \citep{franchini2016lense,zanazzi2019tidal}. Since the X--ray emission is generated primarily in the inner--most accretion region, the disc outer radius will have little effect on the fit results. Finally, the reflection emissivity is set to be $r^{-q}$ within 15$R_g$ and $r^{-3}$ outside 15$R_g$ with $q$ as a free parameter). 

We find that the spectrum XMM\#3 can be well-fit by such a fit--function (C-stat/d.o.f.~$=175.0/159$; Fig.~\ref{fig:xmm3}). This fit statistic is better ($\Delta$AIC~$=8.1$) than that of the phenomenological fit with a power law and a black body. Adding \texttt{relxillCp} into the fit--function improves the fit significantly ($\Delta$AIC~$=96.8$) compared to the fit--function without the \texttt{relxillCp} model component (\texttt{"TBabs*(thcomp*slimdz)"}; C-stat/d.o.f.~$=279.8/163$). We find that the corona is optically thick ($\tau\gtrsim1$) and warm ($kT_e<10$~keV), and the covering fraction of the corona over the disc continuum needs to be less than unity so that $\sim$20\% of the disc photons are observed without being Comptonised. A full list of parameter constraints from the fit is presented in Table~\ref{tb:xmm3par}. Letting the disc's particle density be a free parameter does not improve the goodness--of--fit ($\Delta$AIC~$=-0.1$), and neither does letting the iron abundance be a free parameter ($\Delta$AIC~$=-2$).



\begin{table}
\renewcommand{\arraystretch}{1.5}
\centering
\caption{Best-fit parameter values obtained using a fit--function of \texttt{TBabs*(thcomp*slimdz+relxillCp)} to describe XMM\#3's spectrum. Parameters held fixed, and their values are given in between square brackets. Parameter $f_c$ is the covering fraction of the Comptonising medium. The unit of the normalisation of the \texttt{relxillCp} model is that of flux in the 0.3-10~keV band.}
\begin{tabular}{ccc}
\hline
Model & Parameter & Value \\
\hline
\texttt{TBabs} & $N_H$ / $10^{20}$~cm$^{-2}$ & [1.0]\\
\hline
\texttt{thcomp} & $\tau$ & $8\pm3$\\
& $kT_e$ / keV & $2.4^{+1.4}_{-0.7}$ \\
& $f_c$ & $0.8\pm0.1$ \\
\hline
\texttt{slimdz} & $\dot m$ / $\dot m_{\rm Edd}$ & $>3.3$ \\
 & $\theta$ / $^{\circ}$ & $81^{+4}_{-7}$\\
\hline
\texttt{relxillCp} & $\Gamma$ & $1.5\pm0.2$ \\
& $q$ & $1.6\pm0.8$\\
& log($\xi$) & $0.7\pm0.3$\\
& Refl$_{\rm frac}$ & [$-1$] \\
& log($\rho$ / cm$^{-3}$) & [17] \\
& norm / erg~cm$^{-2}$ s$^{-1}$& $(1.3\pm0.3)\times10^{-5}$\\
\hline
C-stat/d.o.f. & \multicolumn{2}{c}{$175.0/159$}\\
\hline

\end{tabular}
\label{tb:xmm3par}
\end{table}

\subsection{Analysis of UV data using MOSFiT}
\label{sc:uvanalysis}

We analyse \swf{}/UVOT photometry data for the observations listed in Table~\ref{tb:uvobs} using the Modular Open Source Fitter for Transients (MOSFiT; \citealt{guillochon2018mosfit}). The assumptions and details of its TDE module can be found in \citet{mockler2019weighing}. In summary, the module simultaneously fits the UV light curves using a library for model light curves derived using hydrodynamical simulations of TDEs \citep{guillochon2013hydrodynamical}. The fallback mass rate $\dot M_{\rm fb}(t)$ as a function of time is determined from the simulations by varying the $M_{\bullet}$, the stellar mass $M_*$, and the scaled impact parameter $b$\footnote{The impact parameter $\beta$ can be calculated from $b$: if $0\leq~b<1$, $\beta=0.6+1.25b$ when $\gamma=4/3$, and $\beta=0.5+0.4b$ when $\gamma=5/3$; if $1\leq~b\leq2$, $\beta=1.85+2.15(b-1)$ when $\gamma=4/3$, and $\beta= 0.9+1.6(b-1)$ when $\gamma=5/3$. Here $\gamma$ is the polytropic index of the disrupted star so that the equation--of--state of the star is $P\propto\rho^{\gamma}$ ($P$ is the pressure and $\rho$ is the density). See more in \citealt{guillochon2013hydrodynamical} and \citealt{mockler2019weighing}.}. Then the $\dot M_{\rm fb}$ is transformed into a viscously--delayed accretion rate $\dot M_{\rm acc}$ using the viscous timescale $T_{\rm visc}$ (eq.~7 in \citealt{mockler2019weighing}). The module assumes a time--independent efficiency $\epsilon$ in the energy conversion from $\dot M_{\rm acc}c^2$ to a bolometric luminosity $L$ so that $L=\epsilon\dot M_{\rm acc}c^2$. Then assuming thermal radiation, this radiation is emitted from a photosphere with an effective temperature $T_{\rm eff}(L,R_{\rm ph})$, where the photospheric radius $R_{\rm ph}=R_{\rm ph,0}a_{\rm p}(L/L_{\rm Edd})^{l_{\rm ph}}$. Here $R_{\rm ph,0}$ is a normalising factor, and $l_{\rm ph}$ is an exponential index (eq.~9\& 10 in \citealt{mockler2019weighing}). The $a_{\rm p}$ can be regarded as the semimajor axis of the averaged bound orbit of material being accreted when the $\dot M_{\rm fb}$ is at its peak.

The fit parameters are $M_{\bullet}$, $M_*$, $b$, $T_{\rm visc}$, $\epsilon$, $R_{\rm ph,0}$, $l_{\rm ph}$, the difference $t_0$ between the time at peak luminosity and the time of the first detection, and the host column density $N_{\rm H,host}$. Given no intrinsic absorption is found in the X--ray analysis, we fix the $N_{\rm H,host}=1\times10^{19}$~cm$^{-2}$ an order--of--magnitude lower than the Galactic column density in the host direction. No other parameter values are fixed. We use the Markov--Chain--Monte--Carlo (MCMC) routine in MOSFiT to perform the fit. We set 20 walkers and run for 30,000 iterations. We then exclude the first 10,000 iterations as  burn--in. In MOSFiT, the goodness--of--fit and the chain convergence are measured using the Watanabe–Akaike information criteria (WAIC; \citealt{watanabe2010asymptotic}) and the potential scale reduction factor (PSRF; \citealt{gelman1992inference}), where a fit with a PSRF$\le1.2$ is considered to have converged. 

We find the MOSFiT produces a fit to the data with PSRF=1.167 and WAIC=207. A full list of the parameter values and their uncertainties is presented in Table~\ref{tb:uv}. We find the $M_{\bullet}$ is constrained to be between $5\times10^{5}$ and $5\times10^{6}$~M$_\odot$, which is in agreement with our black hole mass measurement using spectral fits to the X-ray data. The systematic errors for each fitting parameter estimated from \citet{mockler2019weighing} are also given in Table~\ref{tb:uv}. We note for parameters other than $M_{\bullet}$, the systematic errors are large compared to the 1$\sigma$ error range given by the fitting procedure. This conclusion also holds if we treat the $M_{\bullet}$ as a known value taken from the analysis of X--ray data (Fig.~\ref{fig:ctxmm}) when performing the fit. We present the data as well as the best--fit MOSFiT model in Fig.~\ref{fig:uvmosfit}.

\begin{table}
\renewcommand{\arraystretch}{1.5}
\centering
\caption{Parameters and their constraints derived from modelling the \swf{}/UVOT light curves using MOSFiT. See the text for the meaning of each parameter. Systematic errors are taken from \citet{mockler2019weighing}.}
\begin{tabular}{ccc}
\hline
Parameter & Value & Systematic Error \\
\hline
log($M_{\bullet}$ / $M_{\odot}$) & $6.2\pm0.3$ & $\pm0.2$ \\
log($M_*$ / $M_{\odot}$) & $-0.01\pm0.09$ & $\pm0.66$ \\
$b$ & $1.2\pm0.2$ & $\pm0.35$ \\
log($T_{\rm visc}$ / days) & $<0.6$ & $\pm0.1$ \\
log($\epsilon$) & $-3.4\pm0.1$ & $\pm0.68$ \\
log($R_{\rm ph,0}$) & $-0.4\pm0.3$ & $\pm0.4$ \\
$l_{\rm ph}$ & $0.05\pm0.04$ & $\pm0.2$ \\
$t_0$ / days & $-19\pm6$ & $\pm15$ \\
\hline
\end{tabular}
\label{tb:uv}
\end{table}

\section{Discussion}
\label{discuss}

In this paper we present \xmm{}/EPIC-pn and \nic{} X-ray spectral analysis of AT2020ocn. The X--ray data can be divided into two periods: an early period with flares (MJD~$\le$59130), and a late period without flares. Over the same period, the UV light curves observed by \swf{} show a gradual decay (Fig.~\ref{fig:lc}). No evidence for UV flares is found. 

We show that the spectra in the early period can be well-fit by a slim disc \citep{wen2020continuum,wen2022library} plus an inverse--Comptonisation (IC) model \citep{zdziarski2020spectral}. Specifically, the spectral evolution along the flares can be explained by variations in the disc inclination and the electron temperature $kT_e$ of the Comptonising medium. Using this fit-function, we constrain the BH mass to be $(7^{+13}_{-3})\times10^{5}$~$M_{\odot}$, while there is no constraint on the BH spin. The best--fit BH mass derived from analysing the UV data with the TDE module of MOSFiT is consistent with this result. Using the empirical $M-\sigma_*$ relation and a velocity dispersion of the stellar absorption lines $\sigma_*=82\pm4$~km/s, \citet{pasham2024lense} estimate the BH mass to be $\sim10^{6.4\pm0.6}$~$M_{\odot}$. The consistency between different methods further prove that AT2020ocn can be explained as a tidal disruption of a star by a massive BH.

The observed inclination variation along the flares can be caused by Lense--Thirring precession during the disc alignment process \citep[e.g.,][]{stone2012observing,franchini2016lense}. Such inclination variations have been observed in simulations \citep[e.g.,][]{fragile2005hydrodynamic,zanazzi2019tidal,white2019tilted}, and they have been proposed to explain the highly--variable jet features in jetted TDEs (e.g., Swift J164449.3+573451; \citealt{tchekhovskoy2014swift,liska2018formation}; see also \citealt{teboul2023unified}). For a fast--spinning BH of $M_{\bullet}=10^{6}$~$M_{\odot}$, the Lense--Thirring precession period during the disc alignment process is calculated to be $\sim10$~days \citep{franchini2016lense,zanazzi2019tidal}, and the timescale for the whole alignment phase is $\lesssim10^{2}$~days \citep{franchini2016lense}. Such timescales for AT2020ocn are in agreement with the calculations (a duration of $\sim 1 - 10$~days for individual flares, and the source stops flaring $\lesssim200$~days after the first \swf{} detection of the event). \citet{pasham2024lense} discovered a $17.0^{+1.2}_{-2.4}$--day quasi--periodicity in \nic{} data in the early--time period, using an energy band of 0.3--1.0~keV that is slightly different from that in our study (0.3--1.1~keV). Without assuming their best-fit period, we fit the early--time \nic{} spectra and find an evolution of the $\theta$ and the IC strength. We investigated if the already--found periodicity in the light curve data (\citealt{pasham2024lense}) can also be found in a Lomb--Scargle periodogram (LSP; \citealt{lomb1976least,scargle1982studies}) of our best-fit $\theta$ or $kT_e$ parameter values as a function of time. An LSP is designed to search for (quasi-)periodic signals in unevenly--sampled time series. In the frequency domain, we find that LSPs of the count rate in 0.3--1.1~keV and the $kT_e$ as a function of time show periodicity peaks that are consistent with those found in \citet{pasham2024lense}, while the LSP of the $\theta$ as a function of time only shows peaks at $\approxlt$3$\sigma$ significance level assuming a white--noise background (Fig.~\ref{fig:lsp}). It is possible that, as the IC component dominates the spectrum when the inclination decreases, most of the periodicity is imprinted in the IC component while the periodicity in $\theta$ is less pronounced. A thorough timing analysis of the \nic{} early--time data, where the impact of the background red--noise is included in estimating the detection significance, can be found in \citet{pasham2024lense}.


\begin{figure*}
    \centering
    \subfloat[]{\includegraphics[width=0.5\textwidth]{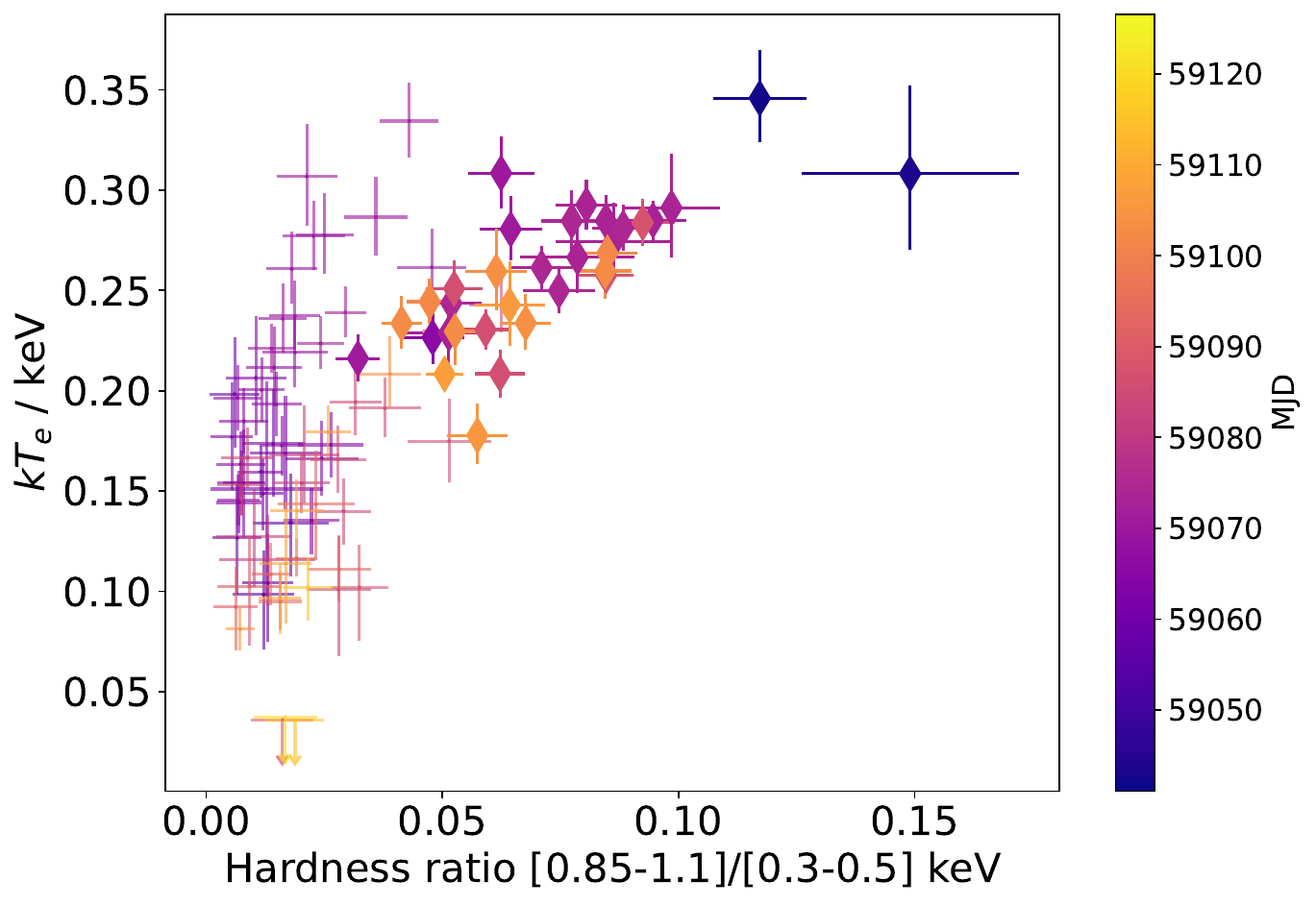}}\hfill
    \subfloat[]{\includegraphics[width=0.5\textwidth]{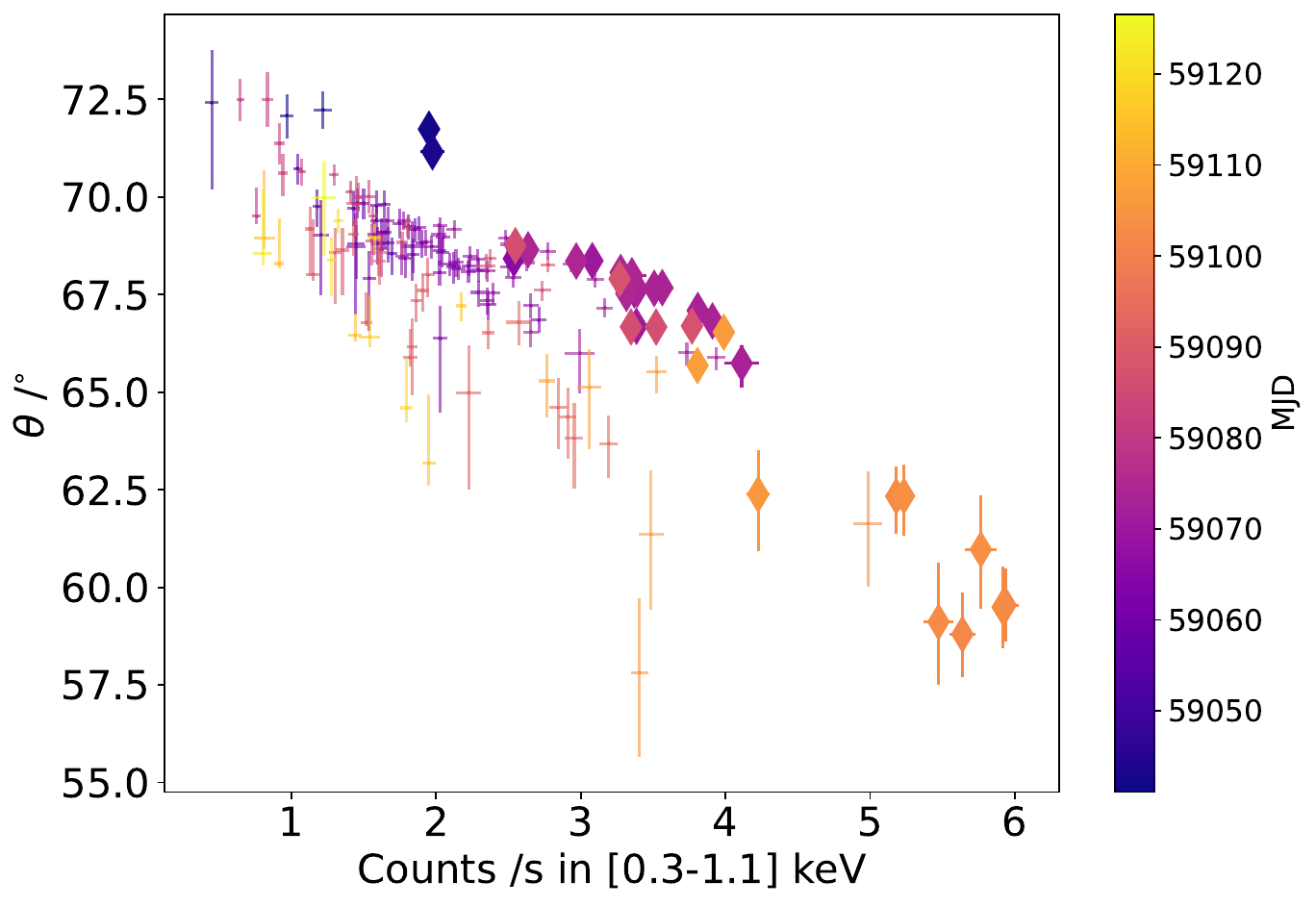}}
    \caption{\textit{Left panel}: The electron temperature $kT_e$ vs.~spectral hardness ratio, derived from the fit procedure producing Fig.~\ref{fig:nicconstraints}. The epochs are colour--coded by their observation time in MJD. Epochs highlighted with diamond markers have a source count rate higher than the background count rate in the 0.85--1.1~keV energy band, so their hardness ratios are least affected by uncertainties in the \nic{} background estimation. The hardness ratio is traced by the temperature $kT_e$ for all flares. \textit{Right panel}: The disc inclination $\theta$ vs.~the total count rate inn the 0.3--1.1~keV energy band. Generally, the lower the $\theta$, the higher the count rate in the 0.3--1.1~keV band. The spectra at the X--ray flare peaks are hard, which can be explained by enhanced inverse--Comptonisation. Some disc photons with energies $<0.3$~keV are up--scattered into the 0.3--1.1~keV band, more than those being scattered out,
    resulting in a higher count rate than that predicted by a model including only a slim disc.}
    \label{fig:pars}
\end{figure*}

The electron temperature $kT_e$ of the IC component traces the spectral hardness ratio ({\it Left panel} Fig.~\ref{fig:pars}): for all flares a higher $kT_e$ is found when the spectrum becomes harder. This correlation is expected: the higher the temperature, the more energy a single photon is likely to get from the up--scattering events before escaping the Comptonising medium, thus the harder the emergent photon spectrum. A possible supply of high--energy electrons for the IC process comes from a disc outflow. For sources at high/super--Eddington accretion rates, a disc outflow is seen in both simulations \citep[e.g.,][]{ohsuga2011global,takeuchi2013clumpy,kitaki2021origins} and observations \citep[e.g.,][]{middleton2013broad,pinto2016resolved,kara2018ultrafast,pinto2021xmm}. However, for
AT2020ocn, the lack of high-resolution X-ray spectral data with a sufficient signal--to--noise ratio (e.g., from \xmm~ RGS) precludes confirmation of the presence of a disc outflow.

In addition, we find an inverse correlation between $\theta$ and the count rate in 0.3--1.1~keV band ({\it Right panel} in Fig.~\ref{fig:pars}). This behaviour can be understood as follows: as the inclination decreases, more photons from the inner disc region are observed. The slim disc model predicts a harder and brighter disc continuum the more the inner disc region is observable \citep[e.g.,][]{wen2020continuum,wen2022library}. As the source flux increases along each flare, we detect the IC component. Some disc photons with energies $<0.3$~keV are up--scattered into the 0.3--1.1~keV band by the Comptonising medium, more than those being scattered out of this energy band, resulting in a higher count rate than that predicted by only a slim disc. We find that the inverse--Comptonisation starts to be detected at higher inclinations along the first flare compared to the subsequent flares, distinguishing the first flare from the rest in both the hardness ratio vs.~count rate diagram (Fig.~\ref{fig:hid} and Fig~.\ref{fig:pars}).

A physical link between the strengthening/weakening of the IC component, and the inclination variation of the disc, is possible, as indicated by the analysis of the early--time \nic{} data. Simulations suggest that when a massive BH accretes at super--Eddington levels, in the scenario of a powerful disc outflow, the temperature of the inner accretion region reaches $\gtrsim10^{6}$~K (corresponding to $\gtrsim0.09$~keV; e.g., \citealt{jiang2019super,yang2023properties}). Our results are consistent with the picture that, as the disc inclination decreases, more photons from the hotter accretion region at smaller radii are observed. Thus, we observe the temperature $kT_e$ of electrons participating the IC process to increase during the X--ray flares, and its value ($\sim0.3$~keV) is similar to the value expected from simulations. 

In our analysis, for the \nic{} spectra that are not well--fit by the slim disc$+$IC model, residuals may be caused by any or severeal of the following reasons: 1) The background in the source$+$background spectrum deviates from the background spectrum generated from \nic{} prescriptions. 2) The source spectrum varies during the exposure of a single epoch (typically $\sim$ks). The slim disc model assumes a steady--state. Given a slim disc of $M_{\bullet}=7\times10^{5}$~$M_{\odot}$, $a_{\bullet}=0.9$, and $\dot m=30\dot m_{\rm Edd}$, the disc viscous timescale $t_{\rm vis}=\alpha^{-1}\Omega^{-1}(R)(H/R)^{-2}$ in the inner disc region (e.g., $\lesssim10$~$R_g$) is $\sim$~ks. Here $\alpha$ is the Shakura--Sunyaev viscosity parameter \citep{shakura1973black}, $H(R)$ is the disc height, and $\Omega(R)$ is the orbital frequency; with such a high $\dot m$, the scaled disc height $H/R$ reaches $\sim0.4$ at $10$~$R_g$. The Photons from the inner disc region dominate the disc spectrum. Thus, on a timescale similar to the exposure time of individual \nic{} epochs, the steady--state approximation of the slim disc model may not be 100\% valid. 3) Meanwhile, other physical mechanisms such as disc outflow \citep[e.g.,][]{middleton2013broad,kara2018ultrafast}, or disc reflection \citep[e.g.,][]{masterson2022evolution} could affect the observed TDE spectrum. With the current data, it is not possible to disentangle different origins of the residuals. Given that most of the early--time \nic{} spectra are describe well by our fit-function, we conclude that a slim disc$+$IC model describes the source X--ray spectrum well, and the observed long--term ($\gtrsim10$~days) variation of the disc inclination is not due to variations in the physics of the disc. 

We test if the enhancement of the IC component during the flares can be explained by an increase in the covering fraction or the optical depth $\tau$ of the Comptonising medium, instead of an increase in $kT_e$. We find that a slim disc plus a Comptonising medium with a varying covering fraction (with fixed $kT_e=0.3$~keV and $\tau=20$) can describe 140 out of 206 early--time \nic{} spectra (C-stat/d.o.f.~$<2$). The value of the covering fraction in a black hole accretion system has sometimes be found to be $\lesssim0.5$ \citep[e.g.,][]{done2006disc,wilkins2015comptonization,chen2022insight,dai2023evolution}, while in other cases it is consistent with the maximum value of 1 \citep[e.g.,][]{tripathi2022thermal,liu2023deciphering,cao2023rapidly}. We re--run the fitting procedure as used for Fig.~\ref{fig:nicconstraints} but fixing the covering fraction to 0.5 instead of 1. In this case, we find that 152 out of 206 spectra are well--fit (Fig.~\ref{fig:nicconstraints-fc0.5}), and both the $\theta$ and $kT_e$ behaviour is similar to Fig.~\ref{fig:nicconstraints}. Similarly, a slim disc plus a Comptonising medium with a varying optical depth (with fixed $kT_e=0.3$~keV and a covering fraction of unity) can describe 153 spectra. It is possible that during each flare multiple parameters of the Comptonising medium are varying as the disc inclination changes. However, with the current data, it is not possible to disentangle the effects of different IC parameters. Nonetheless, in all the tests, the IC becomes prominent at different $\theta$ values between the first and the subsequent flares (e.g., Fig.~\ref{fig:pars}). Our tests on the IC parameters suggest that the physical properties of the Comptonising medium vary between flares.

Given the large uncertainties in constraining the BH spin from the \xmm{} data (Fig.~\ref{fig:ctxmm}), we test if our findings about the $\theta$ and $kT_e$ behaviour are sensitive to the choice of $a_{\bullet}$. We perform the fit procedure described in Section~\ref{sc:nicscan}, but with fixed $M_{\bullet}=5\times10^{5}$~$M_{\odot}$ and $a_{\bullet}=0.2$. We find that 160 spectra are well--fit (Fig.~\ref{fig:nicconstraints-lowa}). The variation behaviour of both $\theta$ and $kT_e$ is qualitatively similar to that in Fig.~\ref{fig:nicconstraints}, but the range over which the inclination varies is much larger for the $a_{\bullet}=0.2$ case (the smallest inclination value within the concerned period from $\sim57^{\circ}$ to near--zero). This enlarged variation is to be expected: for $a_{\bullet}=0.2$ the inner edge of the disc is further away from the BH yielding a lower temperature and subsequently, a softer spectrum, as the innermost--stable--circular--orbit (ISCO) lies further away from the BH event horizon (the inner edge of the disc is set to the ISCO in the {\texttt slimdz}). As the inclination decreases, more hard ($>1$keV) photons from the inner disc are observed, explaining the flares. Therefore, to get the same increase in flux of hard X--ray photons, the inclination must reach a value closer to zero for $a_{\bullet}=0.2$, compared to the situation for $a_{\bullet}=0.9$. Meanwhile, the peak features in $\theta(t)$ with respect to time $t$ occur at similar $t$ for both cases. We conclude that, qualitatively, our findings about the $\theta$ and $kT_e$ behaviour with respect to time are not sensitive to the choice of BH spin during the \nic{} fitting procedure.

We also test if the early--time AT2020ocn data can be equally--well described by a slim disc plus a blackbody component instead of an IC component. We perform the fitting procedure similar to that producing Fig.~\ref{fig:nicconstraints}, but replacing the IC model \texttt{thcomp} by a black--body model \texttt{zbbody}. The \texttt{XSPEC}'s syntax for the test fit--function is then \texttt{TBabs*(zbbody+slimdz)}. We find a large fraction of the \nic{} spectra cannot be well--fitted with a varying black body of $kT_e\sim0.3$~keV (Fig.\ref{fig:nicconstraints-bb}; only 71 out of 206 have a C-stat/d.o.f.~$<2$). Allowing the black body temperature to be free--to--vary during the fitting procedure increases the number of good fits to 115 out of 206 but this is still less than in the number of good fits when using the slim disk plus IC component (see Fig.~\ref{fig:nicconstraints}; 165 out of 206 fits are good, with one more d.o.f.~than the case of \texttt{zbbody} in each individual fit). Therefore, we conclude that a variable IC component provides a better description of the data compared to a varying black body component.
To summarise the different fits and tests we perform for the early--time \nic{} data, we list them in Table~\ref{tb:models}. For reference, we also show the histograms of C-stat/d.o.f.~for each fit in Fig.~\ref{fig:histall}.

\begin{table*}
\renewcommand{\arraystretch}{1.5}
\centering
\caption{Different fit functions used in this paper to fit the \nic{} data in the early--time period. In total, there are 206 spectra to be fit. $kT_{\rm zbb}$ and norm$_{\rm zbb}$ are the temperature and the normalisation of a black body model, respectively. See the text for more details. Histograms of the C-stat/d.o.f.~from each test are presented in Fig.~\ref{fig:histall}}
\begin{tabular}{cccp{7cm}c}
\hline
No. & Model & Free parameter & Fixed parameter & Number of good--fits (C-stat/d.o.f.~$<2$) \\
\hline
1 & \texttt{TBabs*slimdz} & $\dot m$, $\theta$ & $N_{H,G}=1\times10^{20}{\rm cm}^{-2}$, $M_{\bullet}=7\times10^5$~$M_{\odot}$, $a_{\bullet}=0.9$ & 108 \\
\hline
2 & \texttt{TBabs*thcomp*slimdz} & $\dot m$, $kT_e$ & $N_{H,G}=1\times10^{20}{\rm cm}^{-2}$, $M_{\bullet}=7\times10^5$~$M_{\odot}$, $a_{\bullet}=0.9$, $f_c=1$, $\tau=20$, $\theta=74^{\circ}$ & 59 \\
3 & & $\theta$, $kT_e$ & $N_{H,G}=1\times10^{20}{\rm cm}^{-2}$, $M_{\bullet}=7\times10^5$~$M_{\odot}$, $a_{\bullet}=0.9$, $f_c=1$, $\tau=20$, $\dot m=30$~$\dot m_{\rm Edd}$ & 165 \\
4 & & $\theta$, $\tau$ & $N_{H,G}=1\times10^{20}{\rm cm}^{-2}$, $M_{\bullet}=7\times10^5$~$M_{\odot}$, $a_{\bullet}=0.9$, $f_c=1$, $kT_e=0.3$~keV, $\dot m=30$~$\dot m_{\rm Edd}$ & 153 \\
5 & & $\theta$, $kT_e$ & $N_{H,G}=1\times10^{20}{\rm cm}^{-2}$, $M_{\bullet}=7\times10^5$~$M_{\odot}$, $a_{\bullet}=0.9$, $f_c=0.5$, $\tau=20$, $\dot m=30$~$\dot m_{\rm Edd}$ & 152 \\
6 & & $\theta$, $f_c$ & $N_{H,G}=1\times10^{20}{\rm cm}^{-2}$, $M_{\bullet}=7\times10^5$~$M_{\odot}$, $a_{\bullet}=0.9$, $\tau=20$, $kT_e=0.3$~keV, $\dot m=30$~$\dot m_{\rm Edd}$ & 140 \\
7 & & $\theta$, $kT_e$ & $N_{H,G}=1\times10^{20}{\rm cm}^{-2}$, $M_{\bullet}=5\times10^5$~$M_{\odot}$, $a_{\bullet}=0.2$, $f_c=1$, $\tau=20$, $\dot m=30$~$\dot m_{\rm Edd}$ & 160 \\
\hline
8 & \texttt{TBabs*(slimdz+zbbody)} & $\theta$, $kT_{\rm zbb}$, norm$_{\rm zbb}$ & $M_{\bullet}=7\times10^5$~$M_{\odot}$, $a_{\bullet}=0.9$, $\dot m=30$~$\dot m_{\rm Edd}$ & 115 \\
\hline
\end{tabular}
\label{tb:models}
\end{table*}

We note there are several limitations to our modelling. One key assumption in the \texttt{slimdz} model is that the disc is aligned with the BH equatorial plane. 
In this paper we find that the early--time X--ray flares are well--explained by a variable inclination, and we interpret it as results from the disc alignment process. We are not modelling the dynamic solutions of a precessing disc, and there are currently no analytic solutions of a tilted slim disc (i.e., with the disc outside the equatorial plane). Changing $\theta$ in the model is really changing the observer's direction, which is not identical to a precessing disc while the observer is on a fixed position regarding the Kerr metric. Therefore, it is important that we try to estimate the deviations of a disc described by the \texttt{slimdz} model from a tilted disc. From Fig.~\ref{fig:nicconstraints} we find the amplitude of the inclination variation to be $\lesssim10^{\circ}$. In such a case, we estimate a change of $1-\sin^2(90^\circ-10^\circ)\approx3$\% in the tilted disk metric from the approximate equatorial metric of a Kerr black hole, corresponding to a change of $3\%$ in the estimation of accretion rate \citep{page1974disk}. Furthermore, using general relativistic magnetohydrodynamic simulations, \citet{fragile2007global} find that the tilt will change the inner radius of the disc, consequently impacting the efficiency of disk radiation. By studying the disc properties, they conclude that effectively, a tilted disc acts like an un--tilted disc with a lower black hole spin. In this sense, we estimate the tilt--induced X--ray flux change by comparing two slim discs with the same accretion rate but different spin values. Assuming $M_{\bullet}=7\times10^{5}~M_{\odot}$, $\dot m=30~\dot m_{\rm Edd}$, and $\theta=60$, we estimate the flux difference between an un--tilted slim disc with $a_{\bullet}=0.998$, and a 10--degree--tilted disc approximated by a disc with $a_{\bullet}=0.998$sin$(90^{\circ}-10^{\circ})=0.983$, to be $\sim$10\% (from $\sim5.13\times10^{-12}$~erg~cm$^{-2}$~s$^{-1}$ to $\sim4.67\times10^{-12}$~erg~cm$^{-2}$~s$^{-1}$) in the energy range 0.3-1.5~keV. The full impact of tilt on the disc physics necessitates further investigation, which is beyond the scope of this paper.

Meanwhile, parameter constraints ($M_{\bullet}$, $a_{\bullet}, etc.$~) from the \texttt{slimdz} might be impacted by this tilt nature of the disc. We estimate the parameter deviations by comparing the results from the early--time spectra with those from the late--time spectra. XMM\#3 is taken after the flaring period, likely when the disc has settled in the equatorial plane. We re--fit the XMM\#3 spectrum with the model in Table~\ref{tb:xmm3par} but let also $M_{\bullet}$ and $a_{\bullet}$ free--to--vary (though note the consistency issue between \texttt{thcomp} and \texttt{relxillCp}). In this test we find a C-stat/d.o.f.~of 174.4/157 ($\Delta$AIC=-3.4, meaning the improvement of including an additional free parameter is not significant). The BH mass is constrained to be $(1.8^{+1.0}_{-0.5})\times10^6$~$M_{\odot}$, and the spin is constrained to be between -0.62 and 0.97; the inclination is constrained to be $>78^{\circ}$. We conclude that the results from our modelling of the early--time XMM\#1, XMM\#2, and \nic{} data are consistent with those from the late period, and therefore, that the assumption in the \texttt{slimdz} model that the disc lies in the equatorial plane of the BH spin is not the dominant factor in the uncertainty in the parameter estimation.

Furthermore, when using the \texttt{slimdz} model to describe the spectrum emitted by a mis--aligned disc during the alignment process, the best-fit spin value derived from modelling could deviate from the true BH spin, and vary between epochs. We test and find that, due to the data quality, it is not possible to constrain the spin value using the \texttt{slimdz} for individual epochs when we treat the spin as a free parameter in our fits of the early--time \nic{} data. Together with the above tests, and the consistent mass constraint from the $M-\sigma_*$ relation, we conclude that the uncertainties in parameter constraints in our study are dominated by the statistical uncertainty of the data instead of our model assumptions on the early--time spectra with \texttt{slimdz}.

Moreover, we do not consider the possibility of a warped disc in our study. Such a warp could happen during the disc alignment phase \citep[e.g.,][]{franchini2016lense,white2019tilted}, although it is important to realise that the TDE discs are much smaller than AGN and X-ray binary discs (when expressed in Schwarzschild radii). As suggested by pioneering numerical simulations on accretion discs \citep[e.g.,][]{white2019tilted,liska2023radiation}, the disc warp and the disc twist make it possible that, when the inner part of the disc precesses to the phase of minimum inclination (i.e., the inner disc region is face--on), the outer part of the disc is at a different phase so that it intersects our line of sight to the inner disc region and obscures radiations coming from that region. In this sense, the non--planarity of the disc introduces a temporary disc self--obscuration of the inner disc region that will reduce the peak X-ray flux. In such a case, using a planar disc model to fit a warp disc will result in an under--estimation of the amplitude of the inclination variation. However, quantifying the impact of warp on the disc spectrum and the black hole measurements require further studies that incorporate the slim disc and the ray--tracing effects, which we defer to future studies.


From our spectral fits to the X-ray data obtained in the ``late period,'' we find an accretion rate $\dot m>3.3$~$\dot m_{\rm Edd}$ (Table~\ref{tb:xmm3par}), indicating that the source remains at super--Eddington levels throughout the first several hundred days after its detection. This behaviour is consistent with our BH mass constraint
of $M_{\bullet}\sim10^{6}$~$M_{\odot}$. Theory predicts that TDEs arising from such a BH tend to stay at super--Eddington levels for more than hundreds of days \citep{stone2016rates,wen2020continuum}.

An IC component from a corona describes the spectrum well during the late period (Table~\ref{tb:xmm3par}). This corona is not necessarily the same as the Comptonising medium detected in the early period. We do not refer to the one detected in the early period as a "corona" to mark the potential difference. In the late period, the source develops a corona characterised by an optical depth $\tau=8\pm3$ and a $kT_e=2.4^{+1.4}_{-0.7}$~keV. The appearance of a coronal component when the disc accretion rate remains above the Eddington mass accretion rate has also been found in the TDE 3XMM~J150052.0+015452 (J150052; \citealt{cao2023rapidly}). The coronal temperature in AT2020ocn ($2.4^{+1,4}_{-0.7}$~keV) is consistent with that in J150052 ($2.3^{+2,7}_{-0.8}$~keV). The magnetic field in the disc might be critical to power such a corona \citep[e.g.,][]{ghisellini1988synchrotron,merloni2001accretion,veledina2011self,beloborodov2017radiative}. It is possible that the electron temperature of the Comptonising medium in the late period is order--of--magnitude larger than that in the early period. As mentioned earlier, the dominant mechanism powering the Comptonising medium might be different between the early and the late period, since the physical properties of the accretion system (e.g., magnetic field) can be much changed as the disc stops precessing and the accretion rate decreases. In the late period, the disc luminosity decreases as the accretion rate decreases, leading to less cooling of the corona through inverse-Comptonisation, that might also contribute to a hotter corona.

The development of a corona also leads to its emission being reflected from the disc, which is taken into account in our fit in an approximate way by using the \texttt{relxillCp} model. The emissivity index $q$ is constrained to be shallow (<3), which might indicate an extended geometry of the corona or a thick disc \citep[e.g.,][]{mundo2020origin}.  \citet{ezhikode2020correlation} have found that many AGN in Seyfert I galaxies also have a shallow emissivity (fig.~5 in their paper). Therefore, a shallow emissivity index in AT2020ocn in the late period is not without precedent. The fit also suggest a high inclination of the system ($81^{+4}_{-7}$ degree) during the XMM\#3 epoch. This inclination value is consistent with the highest value ($\sim74^{\circ}$) found during the disc alignment phase, within 2$\sigma$ error range (and marginally at 1$\sigma$ error range). After the disc alignment is completed, the disc rests at the high inclination. 

As a comparison, the IC process contributes $<34$~\% of the source flux in the 0.3--1.1~keV band at early epochs, while, at XMM\#3, the photons directly from the corona and from the reflection component contribute $64$~\% of the flux in the 0.3--1.1~keV band. Together with the evidence of the hardness ratio (Fig.~\ref{fig:hid}), we conclude that during the first 500 days after its detection, AT2020ocn transits from a soft, disc--dominated spectral state to a hard, corona--dominated spectral state. Spectral state transitions involving a varying coronal component are also observed in other TDEs \citep[e.g.,][]{bade1996detection,komossa2004huge,wevers2019black,jonker2020implications,wevers2021rapid,cao2023rapidly}. Similarly, many ULXs are believed to be powered by high/super--Eddington accretion onto stellar--mass BHs in X--ray binaries \citep[e.g.,][]{gladstone2009ultraluminous,motta2012discovery,sutton2013ultraluminous}. The strengthening/weakening of a coronal component, when the source accretes at about or at super--Eddington rates, is critical to explain spectral state transitions in ULXs \citep[e.g.,][]{sutton2013ultraluminous,feng2016nature,kaaret2017ultraluminous,barra2022unveiling}. Our findings in AT2020ocn provide further evidence that a varying coronal component might be common in systems accreting at about or at super--Eddington rates.

The UV light curves of AT2020ocn show no flares but instead only a gradual decay (Fig.~\ref{fig:lc}). This is similar to the UV light curve behaviour observed in other TDEs \citep[e.g.,][]{van2020optical}. Assuming a slim disc as observed by \xmm{} (Table~\ref{tb:1and2par}), we estimate the UV flux of the disc to be an order--of--magnitude lower ($\lesssim10^{-29}$~erg/s/cm$^{2}$/Hz with a disc outer radius of 2$R_t\approx117R_g$) than what is observed by \swf{}/UVOT during the X--ray early period (Fig.~\ref{fig:lc}). Similarly, the observed early--time (within several hundred of days after the initial disruption) UV flux in several other TDEs is higher than the expected disc UV flux \citep[e.g.,][]{mummery2020spectral,mummery2023x,wen2023optical}. The decoupling of X-rays and UV (flaring and non-flaring) is also seen in other TDEs \citep[e.g.,][]{2018cowpasham,evans2023monthly,guolo2023x}. It is not possible to distinguish different UV emission mechanisms (e.g., self--intersection shocks in debris streams or a layer reprocessing the X--rays from the inner accretion region) in MOSFiT, which is agnostic about the emission mechanism and only assumes an efficiency $\epsilon$ for the fraction of accretion energy converted to the bolometric luminosity $L=\epsilon\dot M_{\rm acc}c^2$ ($\dot M_{\rm acc}(t)$ is the accretion rate from viscously delayed fall--back mass accretion $\dot M_{\rm fb}(t)$, see also Section~\ref{sc:uvanalysis}). The $\epsilon$ is treated as a time--independent, free parameter in our UV modelling and is found to be $\sim10^{-3}$ (Table~\ref{tb:uv}). The energy emitted by the UV photons is similar to our estimate of the energy dissipated in the circularisation process\footnote{A lower limit on the dissipated energy in the circularisation process is estimated, in an order--of--magnitude way, by comparing the orbital energy of the fallback orbit of the most tightly bound debris (major axis $\sim10^{3}$~$R_g$; \citealt{rees1988tidal}) to the orbital energy of a circularised orbit of radius 2$R_t\approx117R_g$ at the disc outer edge \citep{franchini2016lense}.} ($\gtrsim0.007\dot M_{\rm fb}c^2$). It is possible that the circularisation process plays an important role in powering the early--time UV emission.

Overall, our analysis combining the UV and X--ray data is consistent with the picture where the X--ray flares arise from the inner accretion disc, whose inclination varies during the disc alignment process, while the UV emission comes from another mechanism. Recently, \citet{mummery2024fundamental} developed a method using the late--time UV data to constrain the BH mass in a TDE. For this method to be applicable, it is important that the source has reached a ``plateau phase'' in UV in the late time of the TDE \citep[e.g.,][]{inkenhaag2023late,mummery2024fundamental}, and that the disc emission dominates the detected UV emission. Even though the current UV data is not sufficient to tell if this state is achieved before the end of the observed period, it would be interesting to see if the UV scaling method can provide a mass estimation for the AT2020ocn similar to our result or the $M-\sigma_*$ result using future deep UV observations.

\section{Conclusions}

In this paper, we present our analysis of the X--ray and UV data of TDE AT2020ocn observed by \nic{}, \xmm{}, and \swf{}. The X--ray lightcurve shows strong flares in the first $\approx100$~days, while, over the same period, the UV emission decays gradually. From our X--ray spectral fits using a slim disc model, we constrain the BH mass to be $(7^{+13}_{-3})\times10^{5}$~M$_\odot$ at the 1$\sigma$ (68\%) confidence level. This mass is consistent with that
derived from the analysis of the UV light curve,  log($M_{\bullet}$/$M_{\odot}$)$\sim6.2\pm0.3$, and with that derived from the $M-\sigma_*$ relation, log($M_{\bullet}$/$M_{\odot}$)$\sim6.4\pm0.6$.

We find that the disc alignment process might well be responsible for the qualitatively different behaviour of the X--ray and UV light curves with X--ray flares while the UV emission decays gradually. In particular:

\begin{itemize}

\item The early \xmm{}/EPIC-pn spectra can be well--fit by the slim disc emission. The X--ray flares observed by \nic{} 
can arise from a combination of a varying disc inclination and a varying inverse--Comptonisation component. 

\item We explain the inclination variation in this TDE by proposing that the disc alignment is ongoing, which requires that
the orbital angular momentum vector of the star prior to disruption is mis--aligned from the BH spin angular momentum vector. The inner part of the tilted disc is drawn into the BH equatorial plane due to a combination of the Bardeen--Petterson effect and internal torques. This alignment process causes the inner disc to temporarily precess, explaining the observed inclination variations.

\item The observed spectral variations during the X--ray flares can be explained by the slim disk model convolved with the effect of inverse--Comptonisation. The contribution of the inverse--Comptonisation process to the observed spectrum increases with increasing X-ray photon count rate, 
consistent with observing more up--scattered photons from the inner accretion region as the inclination decreases.

\item The UV light curves for AT2020ocn show no evidence for flares but instead only a gradual decay similar to the UV light curves of other TDEs. 
Most of the UV light likely originates from somewhere other than the accretion disk.
The amount of energy emitted in the UV bands is similar to the estimate of that dissipated in the circularisation process.

\item After the period of X--ray flares, the source spectrum becomes much harder. While the mass accretion rate remains at super--Eddington levels, a corona with an optical depth $\tau=8\pm3$ and an electron temperature $2.4^{+1.4}_{-0.7}$~keV forms after $\sim$300~days. We interpret the late--time \xmm{}/EPIC-pn spectrum as a combination of disc emission, coronal emission, and emission reflected  off the disc. Our findings in AT2020ocn provide further evidence that a varying coronal component might be common in 
systems accreting at about or at super--Eddington rates.

\end{itemize}

\section*{Acknowledgements}

We thank the anonymous referee for insightful comments. This work used the Dutch national e-infrastructure with the support of the SURF Cooperative using grant no.~EINF-3954. This work made use of data supplied by the UK Swift Science Data Centre at the
University of Leicester.
AIZ acknowledges support from NASA ADAP grant \#80NSSC21K0988.

\section*{Data Availability}

All the X--ray data in this paper are publicly available from the HEASARC data archive (https://heasarc.gsfc.nasa.gov/). A reproduction package is available at DOI: 10.5281/zenodo.11162299.



\bibliographystyle{mnras}
\bibliography{references} 



\newpage
\appendix
\section{Supplementary materials}

\clearpage
\onecolumn
\begin{longtable}{cccc}
\caption{\swf{} observations used for the MOSFiT analysis of the UV data in Section~\ref{sc:uvanalysis}.}
\label{tb:uvobs}\\
\hline
\endfirsthead

\multicolumn{4}{c}{Continuation of Table~\ref{tb:uvobs}}\\
\hline
\endhead
\hline
\endfoot
\hline
\multicolumn{4}{c}{End of Table}\\
\endlastfoot
        Observation ID & Date & XRT exposure (s) & UVOT exposure (s) \\ 
        \hline
        00013592001 & 2020-06-25 & 2200 & 2167 \\ 
        00013592002 & 2020-07-02 & 1612 & 1595 \\ 
        00013592003 & 2020-07-09 & 1885 & 1824 \\ 
        00013608001 & 2020-07-11 & 4958 & 4948 \\ 
        00013608002 & 2020-07-11 & 3977 & 3916 \\ 
        00013608003 & 2020-07-13 & 4983 & 4920 \\ 
        00013608005 & 2020-07-15 & 4701 & 4693 \\ 
        00013592004 & 2020-07-16 & 568 & 567 \\ 
        00013608006 & 2020-07-16 & 4483 & 4477 \\ 
        00013608007 & 2020-07-17 & 4603 & 4601 \\ 
        00013608008 & 2020-07-18 & 995 & 987 \\ 
        00013608009 & 2020-07-19 & 241 & 240 \\ 
        00013608010 & 2020-07-20 & 933 & 911 \\ 
        00013592005 & 2020-07-21 & 1402 & 1383 \\ 
        00013608011 & 2020-07-22 & 943 & 929 \\ 
        00013592006 & 2020-07-23 & 2208 & 2175 \\ 
        00013608012 & 2020-07-23 & 1003 & 990 \\ 
        00013608013 & 2020-07-24 & 835 & 823 \\ 
        00013608014 & 2020-07-31 & 963 & 952 \\ 
        00013608016 & 2020-08-04 & 827 & 816 \\ 
        00013608017 & 2020-08-06 & 905 & 895 \\ 
        00013608018 & 2020-08-08 & 938 & 924 \\ 
        00013608019 & 2020-08-10 & 1058 & 1047 \\ 
        00013608020 & 2020-08-20 & 1665 & 1640 \\ 
        00013608021 & 2020-08-22 & 1138 & 1127 \\ 
        00013608022 & 2020-08-24 & 1803 & 1779 \\ 
        00013608023 & 2020-08-26 & 1316 & 1304 \\ 
        00013608024 & 2020-08-28 & 1729 & 1706 \\ 
        00013608025 & 2020-08-30 & 1637 & 1624 \\ 
        00013608026 & 2020-09-01 & 1597 & 0 \\ 
        00013608027 & 2020-09-03 & 2088 & 2054 \\ 
        00013608028 & 2020-09-05 & 2055 & 2003 \\ 
        00013608029 & 2020-09-12 & 1443 & 1432 \\ 
        00013608030 & 2020-09-15 & 915 & 910 \\ 
        00013608031 & 2020-09-18 & 1672 & 1649 \\ 
        00013608032 & 2020-09-21 & 998 & 986 \\ 
        00013608033 & 2020-09-24 & 1382 & 1370 \\ 
        00013608034 & 2020-09-27 & 1363 & 1351 \\ 
        00013608035 & 2020-09-30 & 1413 & 1400 \\ 
        00013608036 & 2020-10-03 & 1387 & 1375 \\ 
        00013608037 & 2020-10-06 & 1038 & 1025 \\ 
        00013608038 & 2020-10-09 & 313 & 303 \\ 
        00013608039 & 2020-10-11 & 1401 & 1368 \\ 
        00013608040 & 2020-10-15 & 1850 & 1827 \\ 
        00013608041 & 2020-10-18 & 1680 & 1668 \\ 
        00013608042 & 2020-10-21 & 268 & 267 \\ 
        00013608043 & 2020-10-23 & 346 & 482 \\ 
        00013608044 & 2020-10-27 & 1256 & 1243 \\ 
        00013608046 & 2020-11-02 & 2174 & 2152 \\ 
        00013608047 & 2020-11-05 & 1951 & 1885 \\ 
        00013608048 & 2020-11-08 & 2136 & 2102 \\ 
        00013608049 & 2020-11-08 & 1988 & 1967 \\ 
        00013608050 & 2020-11-11 & 1842 & 1807 \\ 
        00013608051 & 2020-11-14 & 1434 & 1423 \\ 
        00013608052 & 2020-11-17 & 2023 & 2002 \\ 
        00013608053 & 2020-11-20 & 1908 & 1886 \\ 
        00013608054 & 2020-11-27 & 868 & 856 \\ 
        00013608055 & 2020-11-30 & 835 & 823 \\ 
        00013608056 & 2020-12-03 & 955 & 943 \\ 
        00013608057 & 2020-12-10 & 873 & 860 \\ 
        00013608059 & 2020-12-18 & 880 & 868 \\ 
        00013608060 & 2020-12-21 & 653 & 643 \\ 
        00013608061 & 2020-12-24 & 980 & 969 \\ 
        00013608062 & 2021-02-24 & 707 & 696 \\ 
        00013608063 & 2021-02-27 & 940 & 930 \\ 
        00013608064 & 2021-03-02 & 537 & 531 \\ 
        00013608065 & 2021-03-05 & 955 & 933 \\ 
        00013608066 & 2021-03-11 & 903 & 880 \\ 
        00013608067 & 2021-03-14 & 1063 & 1040 \\ 
        00013608068 & 2021-03-20 & 855 & 843 \\ 
        00013608069 & 2021-03-23 & 1068 & 1056 \\ 
        00013608070 & 2021-05-28 & 68 & 67 \\ 
        00013608071 & 2021-05-31 & 1738 & 1715 \\ 
        00013608072 & 2021-06-06 & 1695 & 1683 \\ 
        00013608073 & 2021-06-18 & 1259 & 1225 \\ 
        00013608074 & 2021-06-21 & 1786 & 1744 \\  

\end{longtable}

\clearpage
\twocolumn

\clearpage
\onecolumn
\begin{longtable}{cccc}
\caption{List of per--GTI--based \nic{} spectra used for the spectral analysis in Section~\ref{sc:nicscan}.}
\label{tb:niceobs}\\
\hline
\endfirsthead

\multicolumn{4}{c}{Continuation of Table~\ref{tb:niceobs}}\\
\hline
\endhead
\hline
\endfoot
\hline
\multicolumn{4}{c}{End of Table}\\
\endlastfoot
        Observation ID & Date & GTI duration (s) & Estimated Source Count Rate (count /s) \\ 
        \hline
3201670101 & 2020-07-11 01:44:46 & 483.0 & $0.97\pm0.05$ \\ 
3201670101 & 2020-07-11 03:17:28 & 373.0 & $1.22\pm0.06$ \\ 
3201670102 & 2020-07-12 04:03:47 & 1717.0 & $1.95\pm0.04$ \\ 
3201670102 & 2020-07-12 21:11:06 & 332.0 & $1.98\pm0.08$ \\ 
3201670103 & 2020-07-13 12:34:07 & 1216.0 & $1.23\pm0.03$ \\ 
3201670108 & 2020-07-18 11:46:49 & 342.0 & $0.45\pm0.05$ \\ 
3201670114 & 2020-07-24 11:47:06 & 1665.0 & $1.18\pm0.03$ \\ 
3201670114 & 2020-07-24 21:04:25 & 1196.0 & $1.43\pm0.04$ \\ 
3201670115 & 2020-07-25 00:10:25 & 1276.0 & $1.46\pm0.04$ \\ 
3201670115 & 2020-07-25 06:22:05 & 1857.0 & $1.04\pm0.03$ \\ 
3201670115 & 2020-07-25 11:00:50 & 1089.0 & $1.56\pm0.04$ \\ 
3201670115 & 2020-07-25 14:06:29 & 1524.0 & $1.69\pm0.04$ \\ 
3201670116 & 2020-07-26 02:29:45 & 1064.0 & $1.50\pm0.04$ \\ 
3201670118 & 2020-07-28 11:47:45 & 1075.0 & $1.64\pm0.04$ \\ 
3201670118 & 2020-07-28 17:59:20 & 1080.0 & $1.59\pm0.04$ \\ 
3201670118 & 2020-07-28 21:04:55 & 526.0 & $1.45\pm0.06$ \\ 
3201670119 & 2020-07-29 09:28:03 & 1097.0 & $1.90\pm0.05$ \\ 
3201670119 & 2020-07-29 12:34:37 & 507.0 & $1.20\pm0.06$ \\ 
3201670119 & 2020-07-29 18:45:43 & 1077.0 & $1.59\pm0.04$ \\ 
3201670119 & 2020-07-29 21:51:45 & 1075.0 & $1.60\pm0.04$ \\ 
3201670120 & 2020-07-30 00:57:23 & 1077.0 & $1.65\pm0.04$ \\ 
3201670120 & 2020-07-30 04:03:28 & 1072.0 & $1.63\pm0.04$ \\ 
3201670120 & 2020-07-30 13:21:03 & 1077.0 & $1.67\pm0.04$ \\ 
3201670120 & 2020-07-30 16:27:03 & 1057.0 & $1.84\pm0.05$ \\ 
3201670120 & 2020-07-30 19:32:43 & 1077.0 & $1.81\pm0.05$ \\ 
3201670121 & 2020-07-31 07:56:45 & 1035.0 & $1.85\pm0.05$ \\ 
3201670121 & 2020-07-31 17:14:26 & 1034.0 & $1.57\pm0.04$ \\ 
3201670121 & 2020-07-31 20:20:26 & 1033.0 & $1.75\pm0.05$ \\ 
3201670121 & 2020-07-31 23:26:07 & 1045.0 & $2.04\pm0.05$ \\ 
3201670122 & 2020-08-01 02:32:07 & 1033.0 & $1.79\pm0.05$ \\ 
3201670122 & 2020-08-01 05:38:07 & 1029.0 & $2.71\pm0.06$ \\ 
3201670122 & 2020-08-01 11:49:49 & 1031.0 & $2.23\pm0.05$ \\ 
3201670122 & 2020-08-01 14:55:43 & 1031.0 & $2.12\pm0.05$ \\ 
3201670122 & 2020-08-01 21:18:12 & 387.4 & $1.84\pm0.08$ \\ 
3201670123 & 2020-08-02 04:52:09 & 1025.0 & $2.02\pm0.05$ \\ 
3201670123 & 2020-08-02 11:03:48 & 1030.0 & $1.54\pm0.04$ \\ 
3201670123 & 2020-08-02 14:09:47 & 1024.0 & $2.02\pm0.05$ \\ 
3201670123 & 2020-08-02 18:48:03 & 1040.0 & $1.97\pm0.05$ \\ 
3201670123 & 2020-08-02 20:21:27 & 1030.0 & $2.29\pm0.05$ \\ 
3201670123 & 2020-08-02 23:27:27 & 1016.0 & $1.88\pm0.05$ \\ 
3201670124 & 2020-08-03 01:00:26 & 397.0 & $1.45\pm0.07$ \\ 
3201670124 & 2020-08-03 05:39:10 & 1027.0 & $2.08\pm0.05$ \\ 
3201670124 & 2020-08-03 08:45:05 & 1024.0 & $2.29\pm0.05$ \\ 
3201670124 & 2020-08-03 14:56:48 & 1029.0 & $2.16\pm0.05$ \\ 
3201670124 & 2020-08-03 18:02:46 & 1017.0 & $2.22\pm0.05$ \\ 
3201670124 & 2020-08-03 22:41:27 & 1033.0 & $2.36\pm0.05$ \\ 
3201670125 & 2020-08-04 04:53:28 & 1012.0 & $2.10\pm0.05$ \\ 
3201670125 & 2020-08-04 07:59:08 & 1032.0 & $1.82\pm0.05$ \\ 
3201670125 & 2020-08-04 11:05:06 & 1006.0 & $1.67\pm0.05$ \\ 
3201670125 & 2020-08-04 14:11:04 & 1023.0 & $2.54\pm0.05$ \\ 
3201670125 & 2020-08-04 17:16:47 & 988.0 & $2.05\pm0.05$ \\ 
3201670125 & 2020-08-04 20:22:47 & 1028.0 & $2.14\pm0.05$ \\ 
3201670125 & 2020-08-04 23:28:26 & 1043.0 & $2.03\pm0.05$ \\ 
3201670126 & 2020-08-05 04:07:26 & 1033.0 & $2.78\pm0.06$ \\ 
3201670126 & 2020-08-05 07:13:26 & 1027.0 & $2.66\pm0.06$ \\ 
3201670126 & 2020-08-05 10:19:06 & 1034.0 & $2.07\pm0.05$ \\ 
3201670126 & 2020-08-05 13:25:07 & 1033.0 & $2.35\pm0.05$ \\ 
3201670126 & 2020-08-05 16:31:06 & 1030.0 & $2.36\pm0.05$ \\ 
3201670126 & 2020-08-05 19:36:47 & 1042.0 & $2.02\pm0.05$ \\ 
3201670126 & 2020-08-05 22:42:46 & 1034.0 & $2.50\pm0.06$ \\ 
3201670127 & 2020-08-06 01:48:47 & 1030.0 & $2.27\pm0.05$ \\ 
3201670127 & 2020-08-06 04:54:27 & 1045.0 & $1.93\pm0.05$ \\ 
3201670127 & 2020-08-06 08:00:26 & 1034.0 & $2.29\pm0.05$ \\ 
3201670127 & 2020-08-06 11:06:20 & 1040.0 & $2.24\pm0.05$ \\ 
3201670127 & 2020-08-06 14:12:06 & 1048.0 & $2.53\pm0.05$ \\ 
3201670127 & 2020-08-06 17:18:08 & 1041.0 & $2.23\pm0.05$ \\ 
3201670127 & 2020-08-06 20:24:01 & 1038.0 & $2.38\pm0.05$ \\ 
3201670127 & 2020-08-06 23:29:47 & 1051.0 & $2.03\pm0.05$ \\ 
3201670128 & 2020-08-07 02:35:51 & 1042.0 & $2.54\pm0.05$ \\ 
3201670128 & 2020-08-07 05:41:27 & 1036.0 & $1.78\pm0.05$ \\ 
3201670128 & 2020-08-07 08:49:41 & 919.0 & $2.64\pm0.06$ \\ 
3201670128 & 2020-08-07 11:53:28 & 1048.0 & $1.76\pm0.05$ \\ 
3201670128 & 2020-08-07 14:59:05 & 1068.0 & $2.39\pm0.05$ \\ 
3201670128 & 2020-08-07 18:05:07 & 1053.0 & $2.77\pm0.06$ \\ 
3201670128 & 2020-08-07 21:10:45 & 1075.0 & $2.13\pm0.05$ \\ 
3201670129 & 2020-08-08 00:16:45 & 1073.0 & $2.48\pm0.05$ \\ 
3201670129 & 2020-08-08 06:28:28 & 1072.0 & $2.50\pm0.05$ \\ 
3201670129 & 2020-08-08 12:40:24 & 1076.0 & $2.63\pm0.05$ \\ 
3201670129 & 2020-08-08 18:52:08 & 1086.0 & $3.10\pm0.06$ \\ 
3201670129 & 2020-08-08 21:57:49 & 1091.0 & $2.93\pm0.06$ \\ 
3201670130 & 2020-08-09 01:03:45 & 1095.0 & $3.08\pm0.06$ \\ 
3201670130 & 2020-08-09 04:09:46 & 1094.0 & $2.94\pm0.06$ \\ 
3201670130 & 2020-08-09 07:16:44 & 1036.0 & $2.90\pm0.06$ \\ 
3201670130 & 2020-08-09 11:54:24 & 1116.0 & $3.37\pm0.06$ \\ 
3201670130 & 2020-08-09 16:33:08 & 1132.0 & $3.27\pm0.06$ \\ 
3201670131 & 2020-08-10 00:17:47 & 1136.0 & $3.17\pm0.06$ \\ 
3201670131 & 2020-08-10 04:56:45 & 1155.0 & $3.39\pm0.06$ \\ 
3201670131 & 2020-08-10 11:08:27 & 1123.0 & $3.94\pm0.06$ \\ 
3201670131 & 2020-08-10 17:20:06 & 1189.0 & $3.73\pm0.06$ \\ 
3201670131 & 2020-08-10 20:26:06 & 1198.0 & $2.66\pm0.05$ \\ 
3201670132 & 2020-08-11 02:53:42 & 328.0 & $2.99\pm0.10$ \\ 
3201670132 & 2020-08-11 05:59:23 & 320.0 & $4.11\pm0.12$ \\ 
3201670132 & 2020-08-11 08:49:27 & 1234.0 & $4.05\pm0.06$ \\ 
3201670132 & 2020-08-11 16:34:06 & 1294.0 & $3.81\pm0.06$ \\ 
3201670133 & 2020-08-12 01:51:47 & 1286.0 & $3.51\pm0.06$ \\ 
3201670133 & 2020-08-12 06:30:28 & 1124.0 & $3.91\pm0.06$ \\ 
3201670133 & 2020-08-12 12:56:02 & 365.0 & $3.39\pm0.10$ \\ 
3201670133 & 2020-08-12 13:02:30 & 363.0 & $3.35\pm0.10$ \\ 
3201670133 & 2020-08-12 16:02:01 & 1317.0 & $3.29\pm0.05$ \\ 
3201670133 & 2020-08-12 18:54:07 & 773.0 & $3.56\pm0.08$ \\ 
3201670133 & 2020-08-12 21:59:48 & 1585.0 & $3.34\pm0.05$ \\ 
3201670133 & 2020-08-12 23:32:45 & 1308.0 & $2.85\pm0.05$ \\ 
3201670134 & 2020-08-13 02:38:45 & 1359.0 & $3.28\pm0.05$ \\ 
3201670134 & 2020-08-13 04:11:45 & 918.0 & $3.32\pm0.07$ \\ 
3201670134 & 2020-08-13 07:17:27 & 1203.0 & $2.97\pm0.05$ \\ 
3201670134 & 2020-08-13 10:23:26 & 1257.0 & $2.64\pm0.05$ \\ 
3201670134 & 2020-08-13 16:35:05 & 1277.0 & $2.39\pm0.05$ \\ 
3201670134 & 2020-08-13 22:46:47 & 1170.0 & $1.62\pm0.04$ \\ 
3201670134 & 2020-08-13 23:06:44 & 496.0 & $1.61\pm0.07$ \\ 
3201670135 & 2020-08-14 18:54:44 & 1436.0 & $0.85\pm0.03$ \\ 
3201670137 & 2020-08-16 09:38:08 & 1375.0 & $0.76\pm0.03$ \\ 
3201670140 & 2020-08-19 05:46:48 & 2083.0 & $0.65\pm0.02$ \\ 
3201670140 & 2020-08-19 10:49:10 & 682.0 & $0.61\pm0.04$ \\ 
3201670141 & 2020-08-20 11:28:57 & 1302.6 & $0.92\pm0.03$ \\ 
3201670141 & 2020-08-20 20:49:25 & 828.0 & $0.83\pm0.04$ \\ 
3201670142 & 2020-08-21 04:14:11 & 1722.0 & $1.41\pm0.03$ \\ 
3201670142 & 2020-08-21 08:55:29 & 1584.0 & $1.07\pm0.03$ \\ 
3201670142 & 2020-08-21 15:04:50 & 1584.0 & $1.46\pm0.04$ \\ 
3201670142 & 2020-08-21 20:01:55 & 842.0 & $1.53\pm0.05$ \\ 
3201670143 & 2020-08-22 00:37:08 & 1372.0 & $0.94\pm0.03$ \\ 
3201670143 & 2020-08-22 05:08:36 & 1804.0 & $1.30\pm0.03$ \\ 
3201670143 & 2020-08-22 14:25:35 & 858.0 & $1.70\pm0.05$ \\ 
3201670144 & 2020-08-22 23:42:54 & 1826.0 & $1.57\pm0.04$ \\ 
3201670144 & 2020-08-23 04:14:46 & 1307.0 & $1.56\pm0.04$ \\ 
3201670144 & 2020-08-23 08:59:32 & 1408.0 & $1.77\pm0.04$ \\ 
3201670144 & 2020-08-23 15:11:55 & 685.0 & $1.45\pm0.07$ \\ 
3201670145 & 2020-08-24 03:35:12 & 1426.0 & $1.81\pm0.04$ \\ 
3201670145 & 2020-08-24 12:52:09 & 772.0 & $2.35\pm0.06$ \\ 
3201670145 & 2020-08-24 17:30:54 & 999.0 & $2.73\pm0.06$ \\ 
3201670145 & 2020-08-24 22:10:32 & 1549.1 & $2.37\pm0.05$ \\ 
3201670146 & 2020-08-25 02:42:49 & 1517.4 & $2.55\pm0.05$ \\ 
3201670146 & 2020-08-25 07:27:32 & 1491.0 & $3.52\pm0.05$ \\ 
3201670146 & 2020-08-25 12:06:08 & 1502.0 & $3.35\pm0.05$ \\ 
3201670146 & 2020-08-25 16:44:07 & 1460.0 & $2.77\pm0.05$ \\ 
3201670146 & 2020-08-25 21:22:49 & 1589.0 & $3.27\pm0.05$ \\ 
3201670147 & 2020-08-26 02:01:48 & 1465.0 & $3.77\pm0.06$ \\ 
3201670147 & 2020-08-26 06:41:29 & 1638.5 & $3.91\pm0.05$ \\ 
3201670147 & 2020-08-26 11:19:46 & 872.0 & $3.70\pm0.07$ \\ 
3201670147 & 2020-08-26 14:26:03 & 1481.0 & $3.32\pm0.05$ \\ 
3201670147 & 2020-08-26 20:39:25 & 1455.0 & $3.79\pm0.06$ \\ 
3201670148 & 2020-08-27 10:38:43 & 1117.0 & $1.94\pm0.05$ \\ 
3201670148 & 2020-08-27 15:11:24 & 1747.0 & $1.43\pm0.04$ \\ 
3201670148 & 2020-08-27 19:50:26 & 1735.3 & $1.13\pm0.03$ \\ 
3201670149 & 2020-08-28 00:46:51 & 679.6 & $1.02\pm0.05$ \\ 
3201670149 & 2020-08-28 02:18:03 & 754.4 & $1.61\pm0.06$ \\ 
3201670149 & 2020-08-28 11:29:08 & 731.9 & $1.15\pm0.05$ \\ 
3201670149 & 2020-08-28 15:57:20 & 260.8 & $1.45\pm0.10$ \\ 
3201670149 & 2020-08-28 16:03:23 & 952.6 & $1.35\pm0.05$ \\ 
3201670149 & 2020-08-28 20:37:03 & 1231.0 & $1.30\pm0.04$ \\ 
3201670150 & 2020-08-29 02:42:46 & 1608.0 & $1.52\pm0.04$ \\ 
3201670150 & 2020-08-29 07:22:25 & 1569.0 & $1.91\pm0.04$ \\ 
3201670150 & 2020-08-29 13:41:27 & 1096.2 & $1.82\pm0.05$ \\ 
3201670150 & 2020-08-29 19:45:21 & 2035.0 & $1.83\pm0.04$ \\ 
3201670151 & 2020-08-30 01:56:06 & 1688.0 & $1.86\pm0.04$ \\ 
3201670151 & 2020-08-30 09:58:40 & 1004.0 & $1.50\pm0.05$ \\ 
3201670151 & 2020-08-30 15:56:11 & 1527.0 & $1.61\pm0.04$ \\ 
3201670152 & 2020-08-31 04:15:17 & 1777.0 & $2.36\pm0.04$ \\ 
3201670152 & 2020-08-31 10:43:41 & 390.0 & $2.57\pm0.09$ \\ 
3201670152 & 2020-08-31 16:39:00 & 901.0 & $2.84\pm0.06$ \\ 
3201670152 & 2020-08-31 16:55:57 & 404.0 & $2.23\pm0.09$ \\ 
3201670152 & 2020-08-31 22:58:00 & 1069.0 & $3.19\pm0.06$ \\ 
3201670153 & 2020-09-01 05:09:58 & 944.0 & $2.96\pm0.06$ \\ 
3201670153 & 2020-09-01 11:15:06 & 1317.0 & $2.91\pm0.06$ \\ 
3201670154 & 2020-09-01 23:43:31 & 974.0 & $3.28\pm0.07$ \\ 
3201670155 & 2020-09-09 15:35:28 & 772.0 & $5.94\pm0.09$ \\ 
3201670155 & 2020-09-09 21:47:27 & 953.0 & $5.23\pm0.08$ \\ 
3201670156 & 2020-09-10 03:57:28 & 777.0 & $5.64\pm0.09$ \\ 
3201670156 & 2020-09-10 10:12:10 & 1149.0 & $5.18\pm0.07$ \\ 
3201670156 & 2020-09-10 16:36:04 & 576.0 & $5.47\pm0.10$ \\ 
3201670156 & 2020-09-10 22:39:37 & 1023.0 & $5.92\pm0.08$ \\ 
3201670157 & 2020-09-11 04:56:23 & 577.0 & $4.98\pm0.10$ \\ 
3201670157 & 2020-09-11 10:54:43 & 517.0 & $5.77\pm0.11$ \\ 
3201670157 & 2020-09-11 23:18:17 & 983.0 & $5.23\pm0.07$ \\ 
3201670158 & 2020-09-12 05:30:23 & 957.0 & $4.67\pm0.07$ \\ 
3201670158 & 2020-09-12 11:41:35 & 1025.0 & $3.41\pm0.06$ \\ 
3201670158 & 2020-09-12 17:51:33 & 715.0 & $3.82\pm0.08$ \\ 
3201670159 & 2020-09-13 06:11:07 & 713.0 & $4.23\pm0.08$ \\ 
3201670160 & 2020-09-14 13:28:31 & 869.0 & $3.99\pm0.07$ \\ 
3201670160 & 2020-09-14 19:28:45 & 518.0 & $3.65\pm0.09$ \\ 
3201670161 & 2020-09-15 01:31:07 & 2153.0 & $3.81\pm0.04$ \\ 
3201670161 & 2020-09-15 07:43:54 & 856.0 & $3.53\pm0.07$ \\ 
3201670161 & 2020-09-15 20:10:18 & 515.0 & $3.06\pm0.08$ \\ 
3201670162 & 2020-09-16 08:36:46 & 532.0 & $3.48\pm0.09$ \\ 
3201670162 & 2020-09-16 14:50:07 & 1576.0 & $2.92\pm0.05$ \\ 
3201670162 & 2020-09-16 21:05:09 & 934.0 & $2.77\pm0.06$ \\ 
3201670167 & 2020-09-21 04:44:27 & 271.0 & $0.81\pm0.07$ \\ 
3201670167 & 2020-09-21 17:06:48 & 1013.0 & $1.44\pm0.05$ \\ 
3201670168 & 2020-09-22 11:51:05 & 1098.0 & $1.95\pm0.05$ \\ 
3201670168 & 2020-09-22 17:51:57 & 856.0 & $2.52\pm0.06$ \\ 
3201670169 & 2020-09-23 04:48:04 & 1333.0 & $0.92\pm0.03$ \\ 
3201670169 & 2020-09-23 06:20:43 & 1440.0 & $1.53\pm0.04$ \\ 
3201670170 & 2020-09-24 03:47:48 & 2315.0 & $2.18\pm0.04$ \\ 
3201670170 & 2020-09-24 17:51:47 & 1876.0 & $1.08\pm0.03$ \\ 
3201670171 & 2020-09-25 00:08:10 & 427.0 & $1.54\pm0.07$ \\ 
3201670171 & 2020-09-25 20:04:12 & 1103.0 & $1.57\pm0.05$ \\ 
3201670172 & 2020-09-26 08:27:44 & 1140.0 & $0.74\pm0.04$ \\ 
3201670172 & 2020-09-26 21:09:11 & 1249.0 & $1.80\pm0.05$ \\ 
3201670174 & 2020-09-28 14:45:24 & 1024.0 & $2.64\pm0.06$ \\ 
3201670175 & 2020-09-29 03:03:50 & 2083.0 & $1.32\pm0.03$ \\ 
3201670175 & 2020-09-29 15:33:30 & 2154.0 & $1.28\pm0.03$ \\ 
3201670176 & 2020-09-30 22:31:45 & 334.0 & $0.80\pm0.07$ \\ 
3201670177 & 2020-10-01 10:55:46 & 1453.0 & $1.01\pm0.04$ \\ 
3201670177 & 2020-10-01 17:01:30 & 1599.0 & $0.93\pm0.03$ \\ 
3201670179 & 2020-10-03 06:37:27 & 354.4 & $0.71\pm0.06$ \\ 
3201670180 & 2020-10-04 13:19:08 & 303.9 & $1.23\pm0.08$ \\ 
3201670184 & 2020-10-08 02:45:23 & 1117.0 & $1.11\pm0.04$ \\

\end{longtable}

\clearpage
\twocolumn

\begin{figure*}
    \centering
    \includegraphics[width=\linewidth]{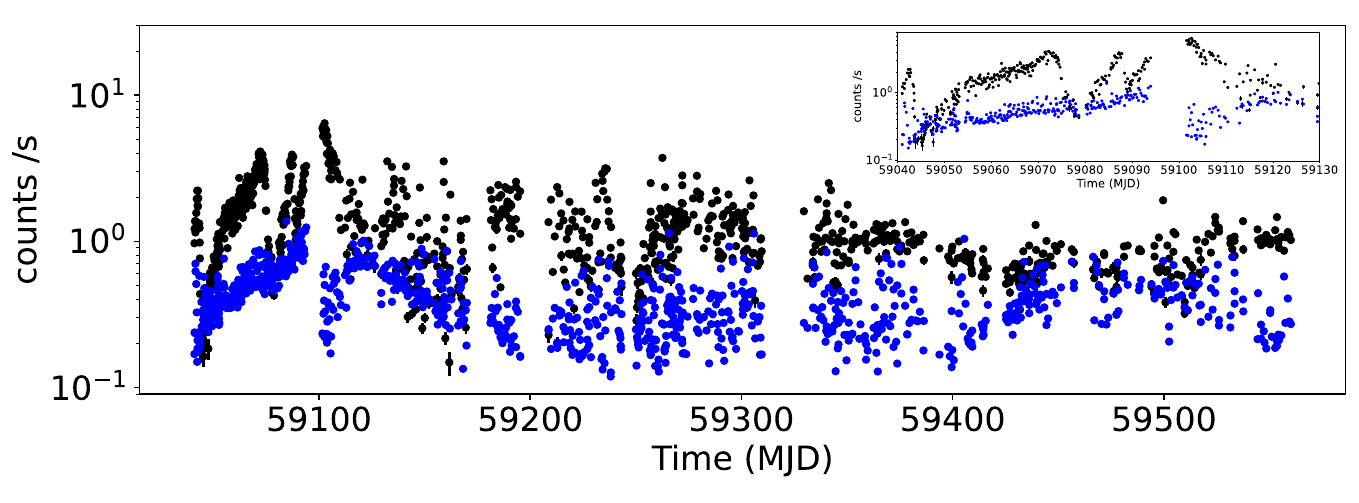}
    \caption{The \nic{} light curve of the 0.3-1.1 keV band (black) and the background light curve of the 0.3-1.1 keV band (blue). The insert shows the zoomed--in view of the early flares (MJD$<59130$).}
    \label{fig:nicbkg}
\end{figure*}

\begin{figure*}
    \centering
    \includegraphics[width=\linewidth]{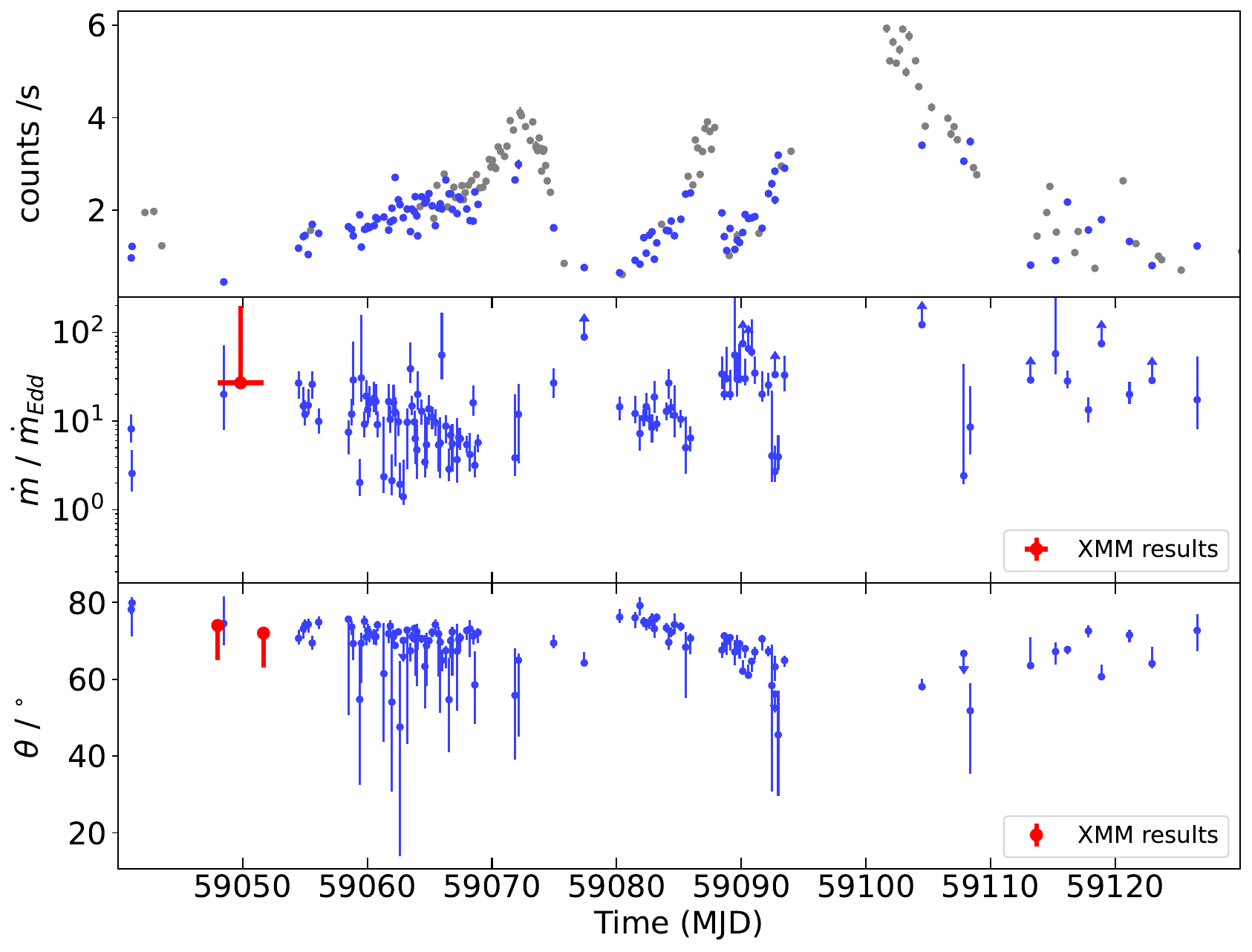}
    \caption{\textit{Top panel}: The \nic{} 0.3-1.1~keV early--time source light curve. We only consider the epochs where the source flux stays above the background level from 0.3~keV to at least 0.6~keV (a total of 206 epochs in the early period). Spectra in epochs marked by blue dots are well--fit (C-stat/d.o.f.~$<2$; 108 in total) by the model in this analysis, while grey dots mark spectra that have C-stat/d.o.f.~$>2$. The fit--function here is different from Fig.~\ref{fig:nicconstraints}: \texttt{TBabs*slimdz}.
    \textit{Middle panel}: Constraints on the disc mass accretion rate $\dot m$ derived from the \nic{} spectra obtained before MJD~59130. We fit each of the 206 spectra individually allowing the accretion rate $\dot m$ and the inclination $\theta$ to vary. We use $N_H=1\times10^{20}$~cm$^{-2}$ for all spectral fits in this paper. We fix $M_{\bullet}=7\times10^{5}$~$M_{\odot}$ and $a_{\bullet}=0.9$ based on the results of the spectral fits to the \xmm{} data (Fig.~\ref{fig:ctxmm}). We show the fitted parameter values for the 108 out of the 206 spectra where the C-stat/d.o.f.~$<2$ (the results from the blue points in the top panel). The accretion rate constrained from the joint--fit to the XMM\#1 and XMM\#2 X--ray spectra is marked with the red dots.
    \textit{Bottom panel}: Constraints on the inclination $\theta$ derived from the same fitting procedure described above.
    We conclude that the slim disc alone cannot explain the early--time \nic{} data well, especially the hard spectra at the peak of the flares are not well fit.}
    \label{fig:nicslim}
\end{figure*}

\begin{figure*}
    \centering
    \includegraphics[width=0.5\textwidth]{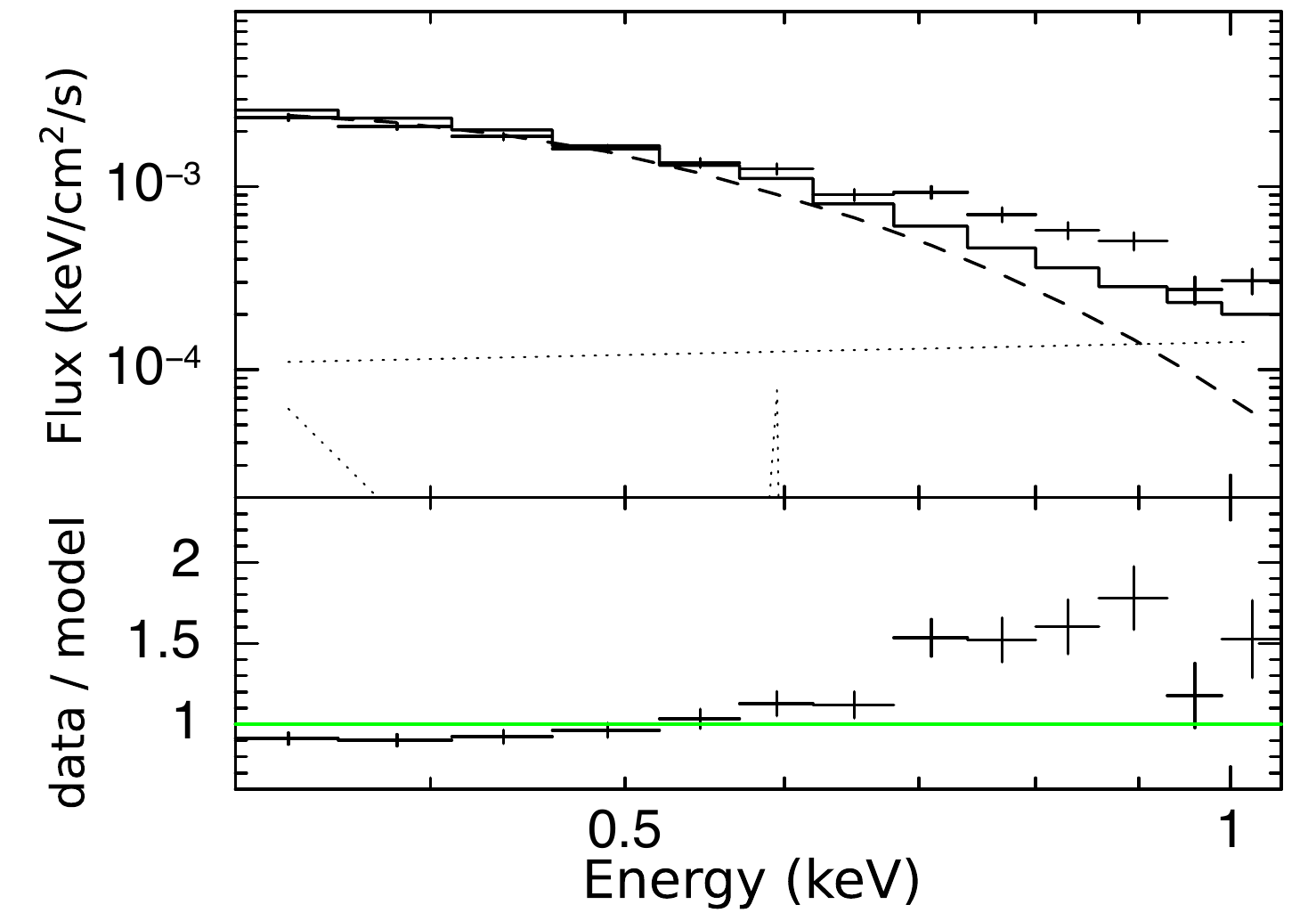}
    \caption{Top panel: An example of the hard \nic{} spectra at flare peaks, taken on MJD~59074.4. The fit--function is comprised of the following model components: \texttt{TBabs*slimdz}. The fit procedure is described in Section~\ref{sc:nicscan} and Fig.~\ref{fig:nicslim}. The solid, dashed, and dotted lines represent the total model, the slim disc emission, and the contribution from the background as determined from fitting the estimated background--only spectrum separately, respectively. The best--fit background power--law indices and Gaussian parameters have been held constant during the fit to the source$+$background spectra. Bottom panel: We show the ratio between the observed number of counts (data) and the predicted number of counts in each bin (model). The spectra at flare peaks are much harder than a continuum described by a slim disc.}
    \label{fig:nicexamp}
\end{figure*}

\begin{figure*}
    \centering
    \subfloat[]{\includegraphics[width=0.5\textwidth]{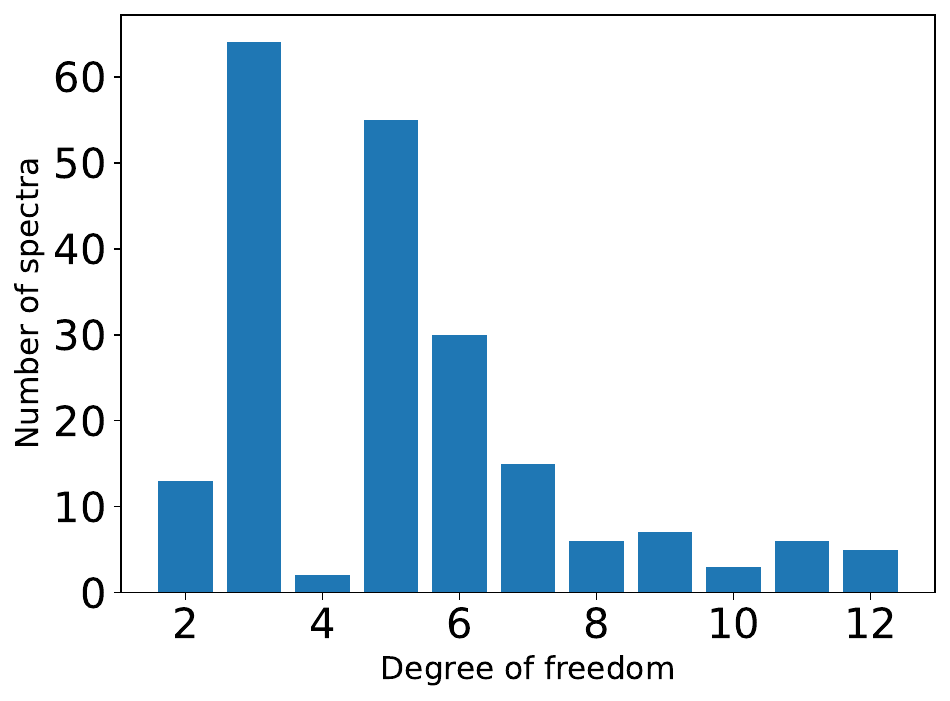}}\hfill
    \subfloat[]{\includegraphics[width=0.5\textwidth]{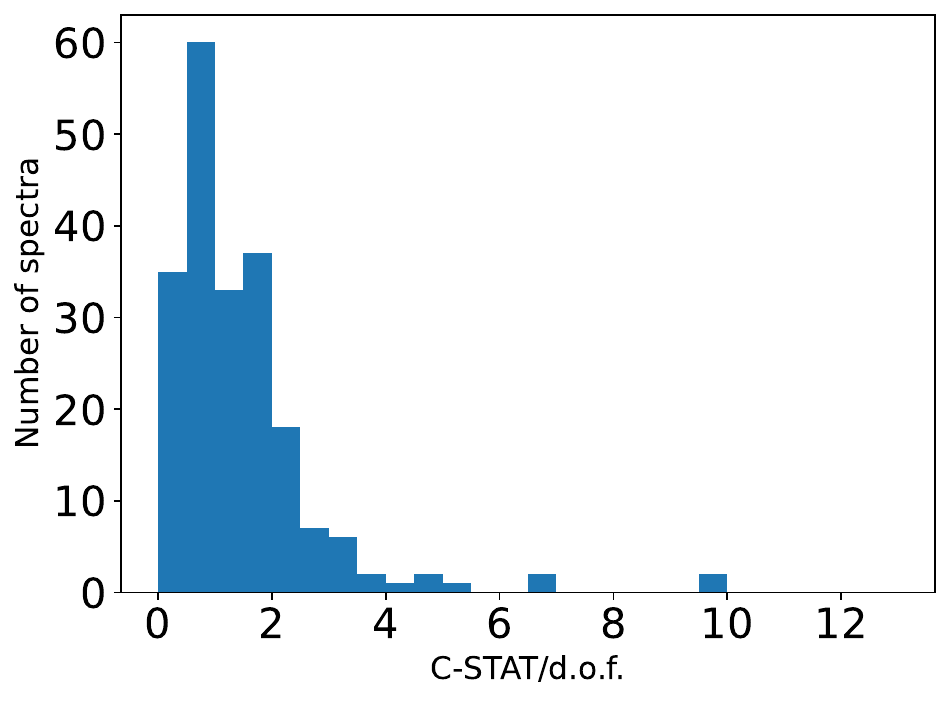}\label{fig:4a}}
    \caption{\textit{Left}: Histogram of the degree--of--freedom (d.o.f) in each fit produced by the fit procedure described in Section~\ref{sc:nicscan} and of which the results are shown in Fig.~\ref{fig:nicconstraints}. Each spectrum has at least 2~d.o.f.~left to be fitted with the 2--parameter ($\theta$ and $kT_e$) fit function. \textit{Right}: Histogram of the C-stat/d.o.f.~produced by the same fit procedure.}
    \label{fig:stathist}
\end{figure*}

\begin{figure*}
    \centering
    \includegraphics[width=0.5\textwidth]{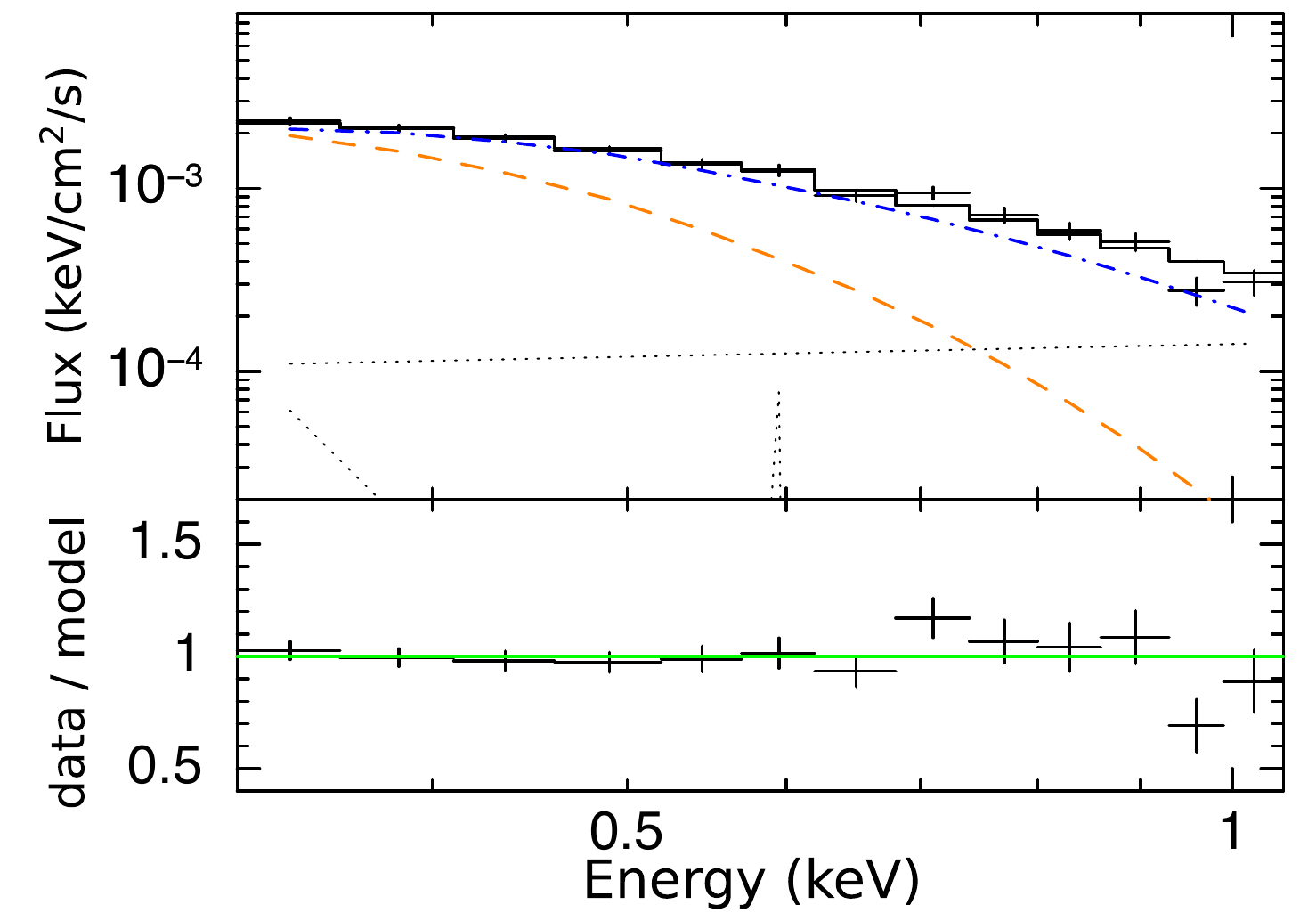}
    \caption{Same as Fig~\ref{fig:nicexamp} but with a fit--function comprising of the following model components: \texttt{TBabs*thcomp*slimdz}. Moreover, the orange dashed line represents the slim disc emission before the inverse--Comptonisation, and the blue dash--dotted line represents the slim disc emission after the inverse--Comptonisation. The C-stat/d.o.f.~$=13.9/11$.}
    \label{fig:nicesoftexamp}
\end{figure*}

\begin{figure*}
    \centering
    \includegraphics[width=0.5\textwidth]{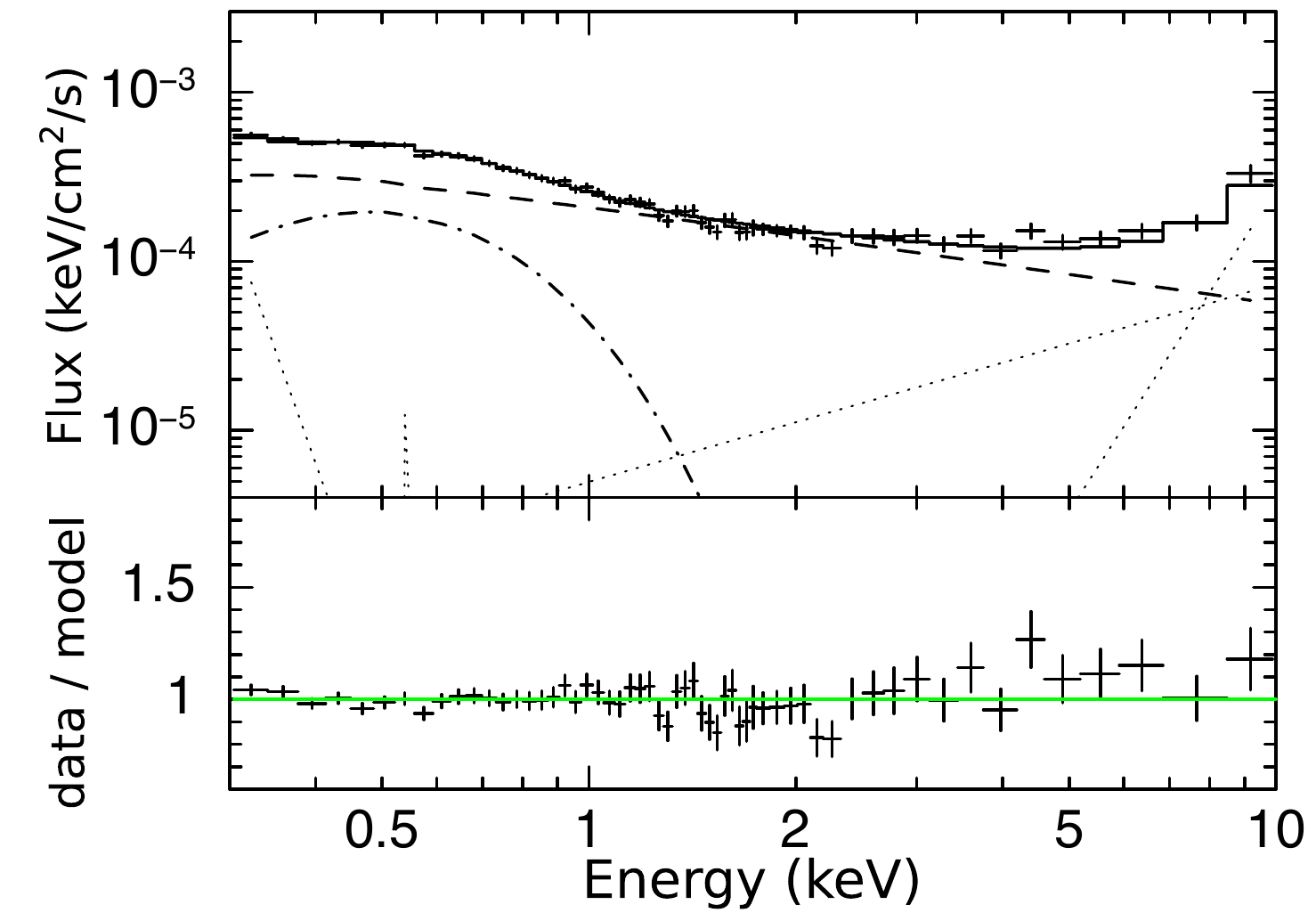}
    \caption{Top panel: phenomenologically fitting the XMM\#3 spectrum with a fit function comprised of a power--law and a black body. In \texttt{XSPEC}'s syntax, the fit function is \texttt{"TBabs*(powerlaw+zbbody)"}. The sold, dashed, dot--dashed, and dot lines represent the best--fit total model, the power--law, the black body emission, and the contribution from the background as determined from fitting extracted spectra from background--only data separately, respectively. Bottom panel: The ratio between the observed number of counts (data) and the best--fit predicted number of counts in each spectral bin (model).}
    \label{fig:xmm3phe}
\end{figure*}

\begin{figure*}
    \centering
    \includegraphics[width=0.8\textwidth]{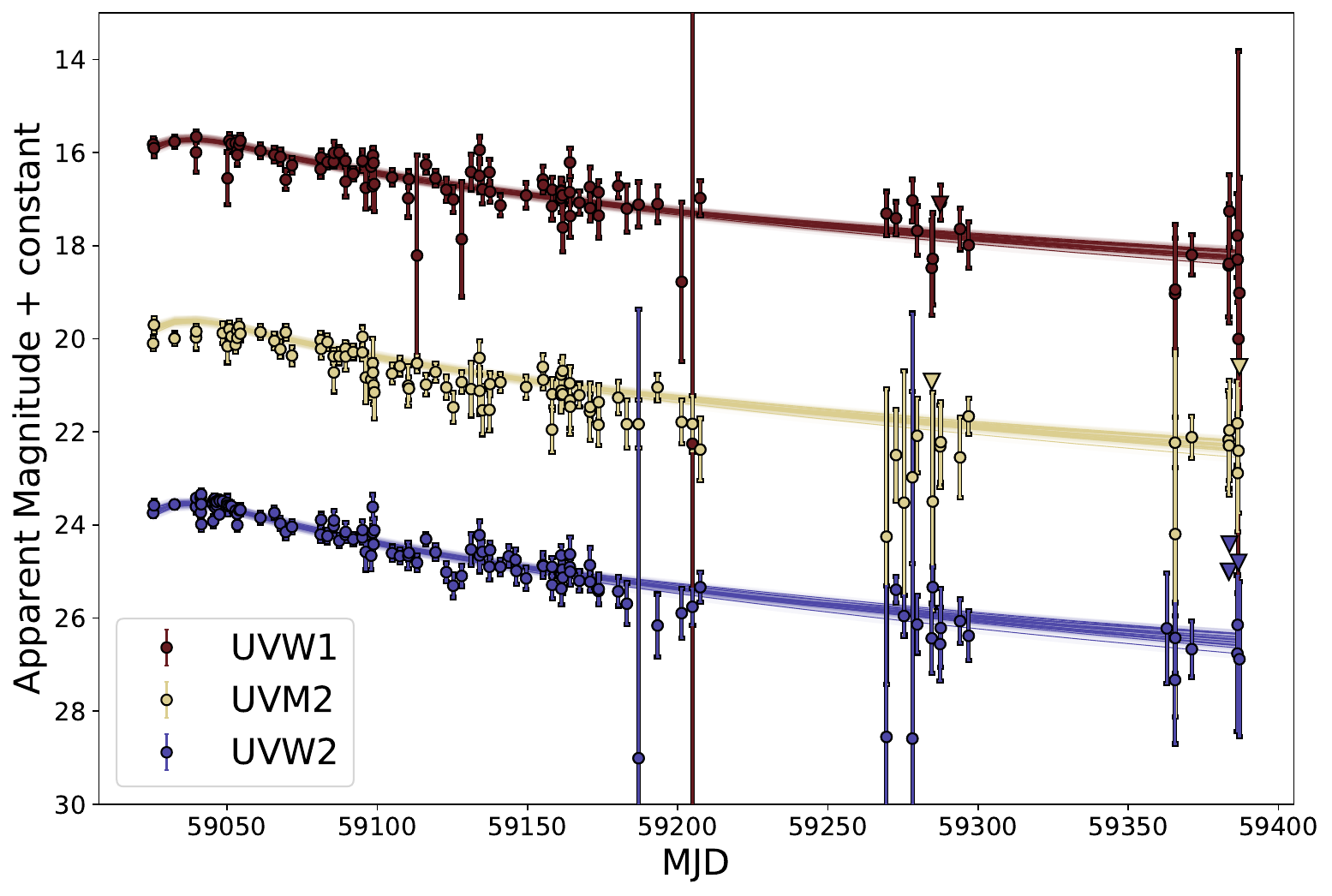}
    \caption{\swf{}/UVOT data fitted using the MOSFiT, as described in Section \ref{sc:uvanalysis}. We shift the y--axis of UVW1 and UVW2 bands by a constant (-4 for UVW1 and +4 for UVW2) for plot clarity.}
    \label{fig:uvmosfit}
\end{figure*}

\begin{figure*}
    \centering
    \subfloat[]{\includegraphics[width=0.9\textwidth]{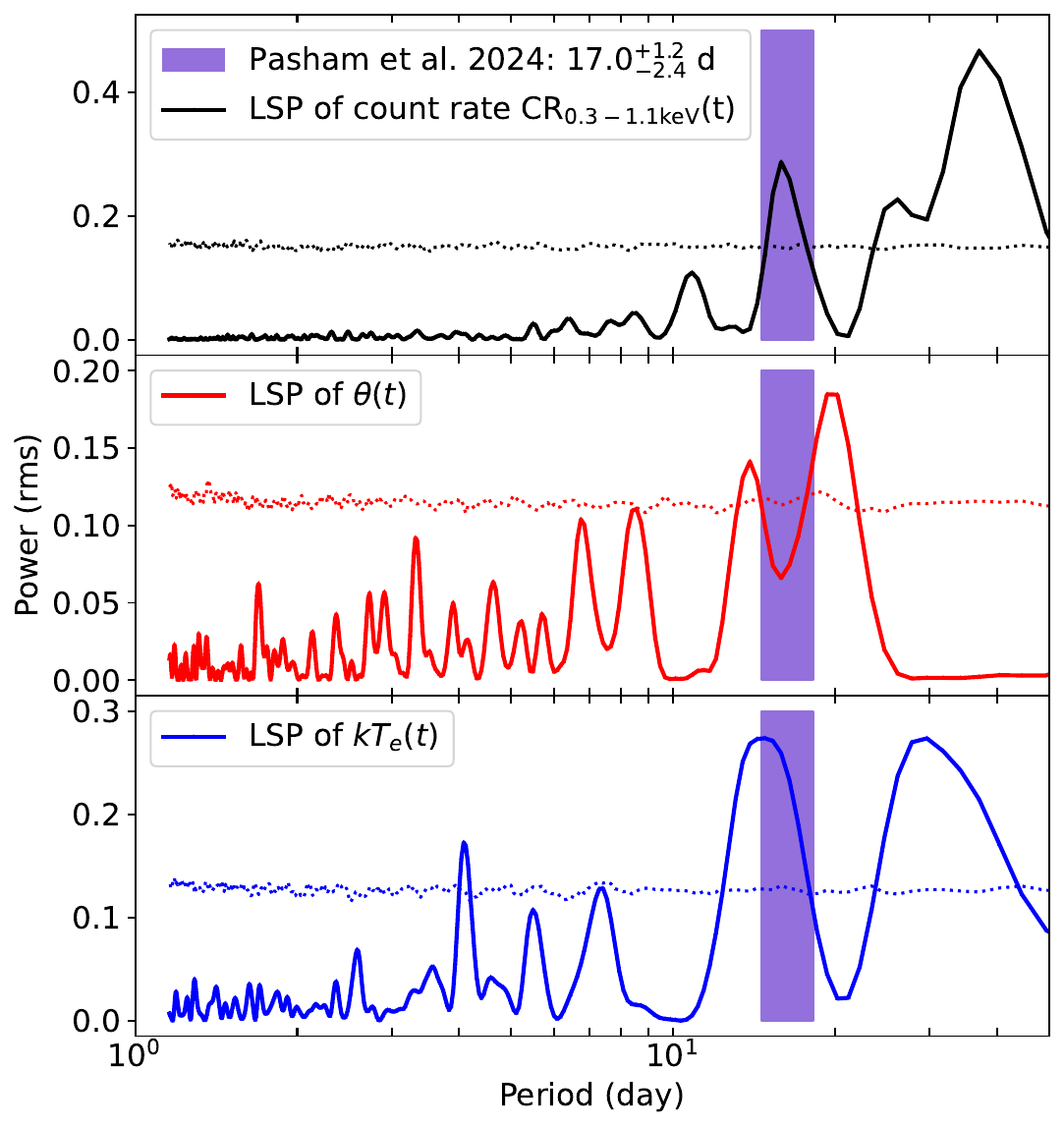}}
    \caption{
    Lomb--Scargle periodograms (LSPs) calculated from the X--ray count rate (CR) in 0.3--1.1~keV (\textit{solid black}), the measured inclination $\theta$ (\textit{solid red}) and electron temperature $kT_e$ (\textit{solid blue}), from the results in Fig.~\ref{fig:nicconstraints}. The x--axis is the period, and the y-axis is the LSP power for a given periodic mode. All three parameters are derived or measured over the period MJD~59041--59130. Dotted lines are the 3$\sigma$ detection thresholds, which is roughly estimated by a bootstrapping approach similar to that in \citealt{evans2023monthly}: for each time series (the count rate, $\theta$, or $kT_e$), we redistribute the value of the parameter under consideration randomly among the same set of time bins as the original, simulating a new time series. We perform 10000 such simulations and then calculate the Lomb--Scargle periodogram of each simulation. The 99.7 percentile of power at each frequency is then calculated. We note a better treatment of the significance estimation should include modelling the red noise in the lightcurves, which is beyond the purpose of this paper and can be found in \citealt{pasham2024lense}. The periodicity peaks in the count rate and $kT_e$ are consistent with a period of $17^{+1.2}_{-2.4}$~days (\textit{purple}) found by \citealt{pasham2024lense}, while the LSP of $\theta$ only shows two weak peaks around 17~days. As the IC component dominates the spectrum during the flares, it is possible that most of the periodicity is imprinted in the IC component, while the periodicity in $\theta$ is less observed. A double--period peak can be found in the LSPs of the count rate, and the $kT_e$. Meanwhile, the LSP of $kT_e$ also shows a 4--day peak that barely reaches the 3$\sigma$ level, which corresponds to no signals in the data (the LSP of the 0.3-1.1~keV count rate lightcurve) and is likely a noise component.
    }
    \label{fig:lsp}
\end{figure*}

\begin{figure*}
    \centering
    \includegraphics[width=\linewidth]{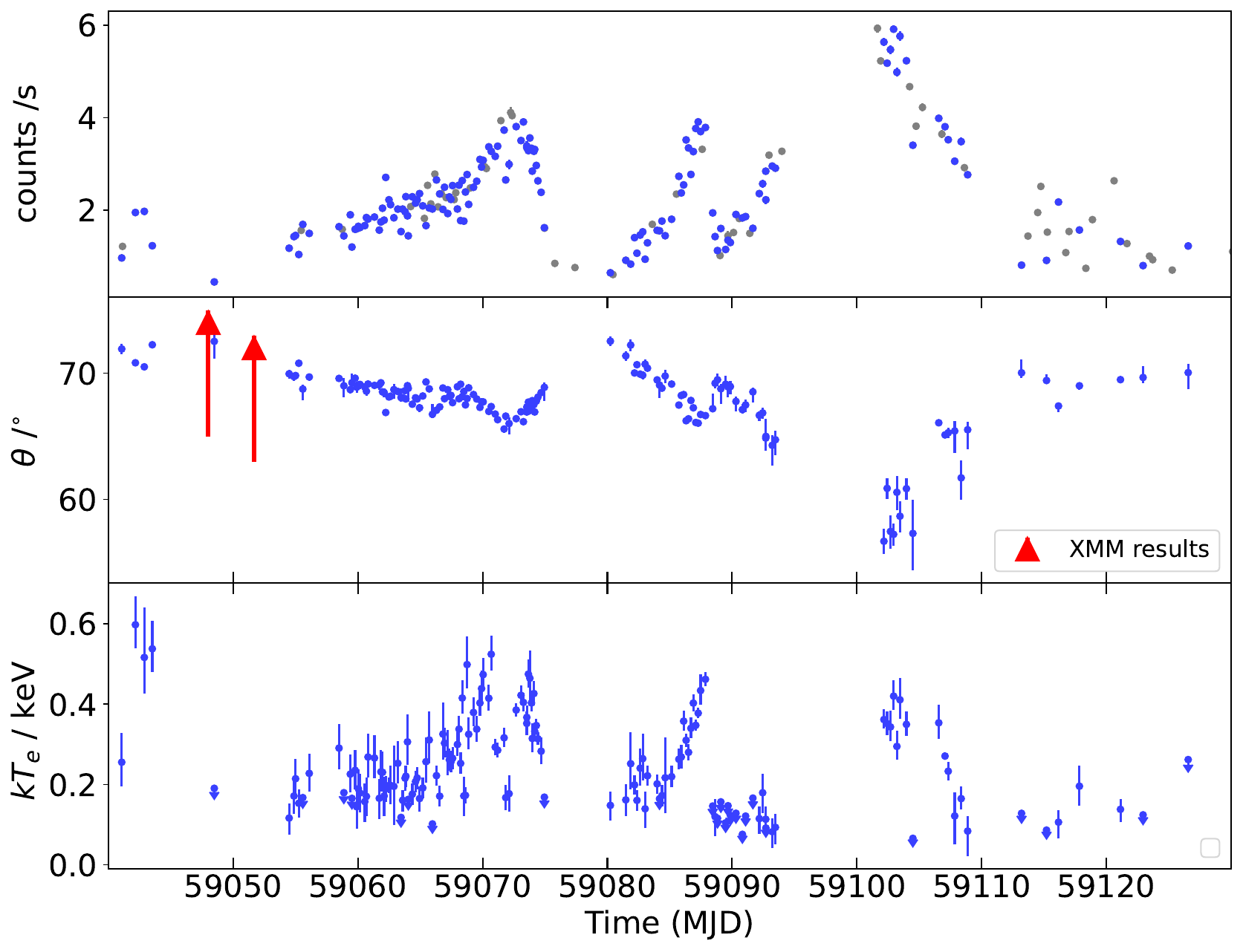}
    \caption{Same as Fig.~\ref{fig:nicconstraints}, but in deriving these panels, we fixed the covering fraction of the IC component to 0.5 instead of unity when fitting the \nic{} spectra. In total, 152 out of 206 spectra are fitted with C-stat/d.o.f.~$<2$ (blue data points).}
    \label{fig:nicconstraints-fc0.5}
\end{figure*}

\begin{figure*}
    \centering
    \includegraphics[width=\linewidth]{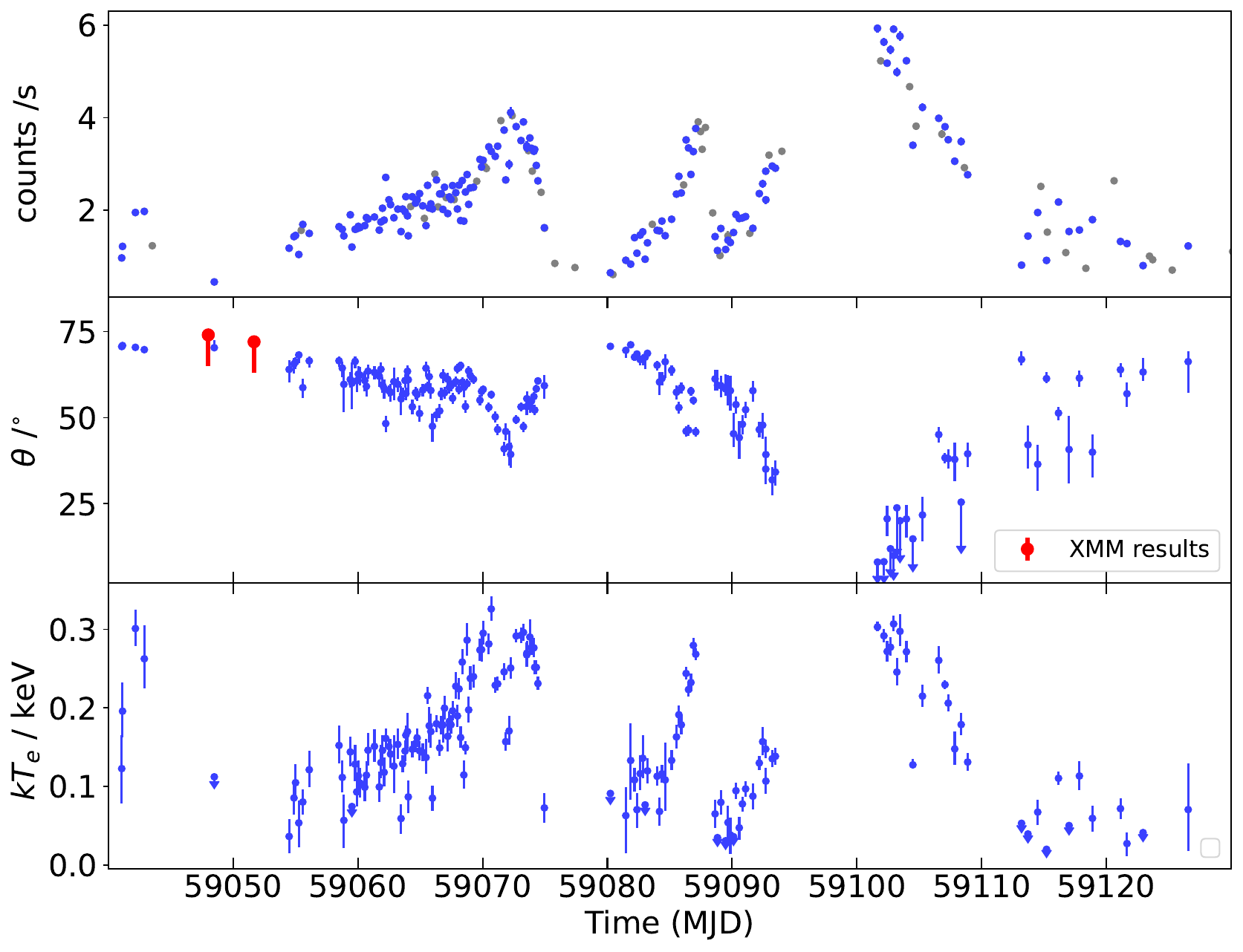}
    \caption{Same as Fig.~\ref{fig:nicconstraints}, but we fix $M_{\bullet}$ to $5\times10^5$~$M_{\odot}$ and $a_{\bullet}$ to $0.2$ when fitting the \nic{} spectra. In total, 160 out of 206 spectra are fitted with C-stat/d.o.f.~$<2$.}
    \label{fig:nicconstraints-lowa}
\end{figure*}

\begin{figure*}
    \centering
    \includegraphics[width=\linewidth]{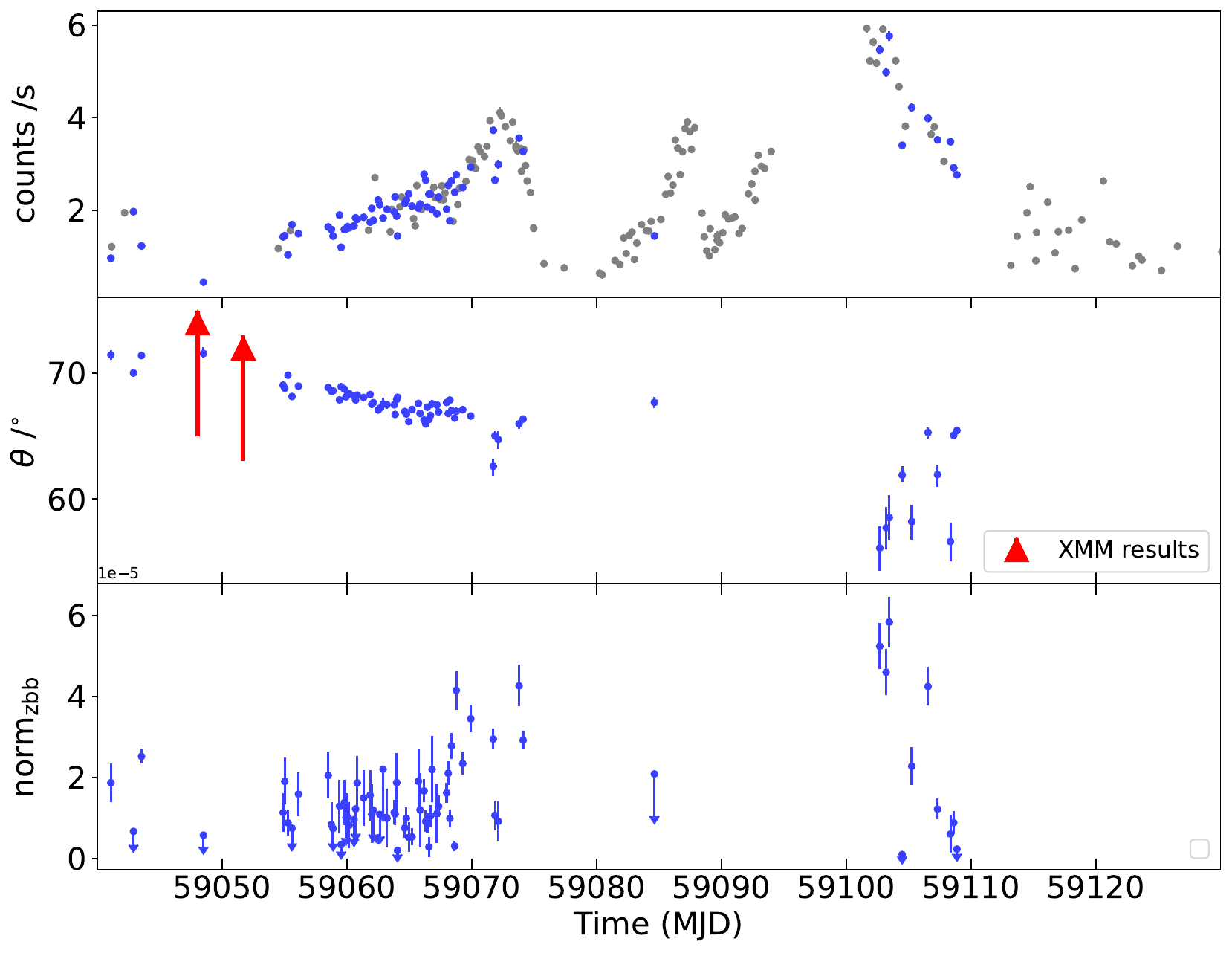}
    \caption{Same as Fig.~\ref{fig:nicconstraints}, but to derive these panels we used a fit--function of \texttt{TBabs*(slimdz+zbbody)} when fitting the \nic{} spectra. The black body temperature is fixed to 0.3~keV during the fit. In total, 71 out of 206 spectra are fitted with C-stat/d.o.f.~$<2$ (blue data points in the \textit{top} panel). Allowing the black body temperature to be free--to--vary during the fit would increase the number of good--fits to 115 out of 206 but still far less than Fig.~\ref{fig:nicconstraints} (165 out of 206).}
    \label{fig:nicconstraints-bb}
\end{figure*}

\begin{figure*}
    \centering
    \includegraphics[width=\linewidth]{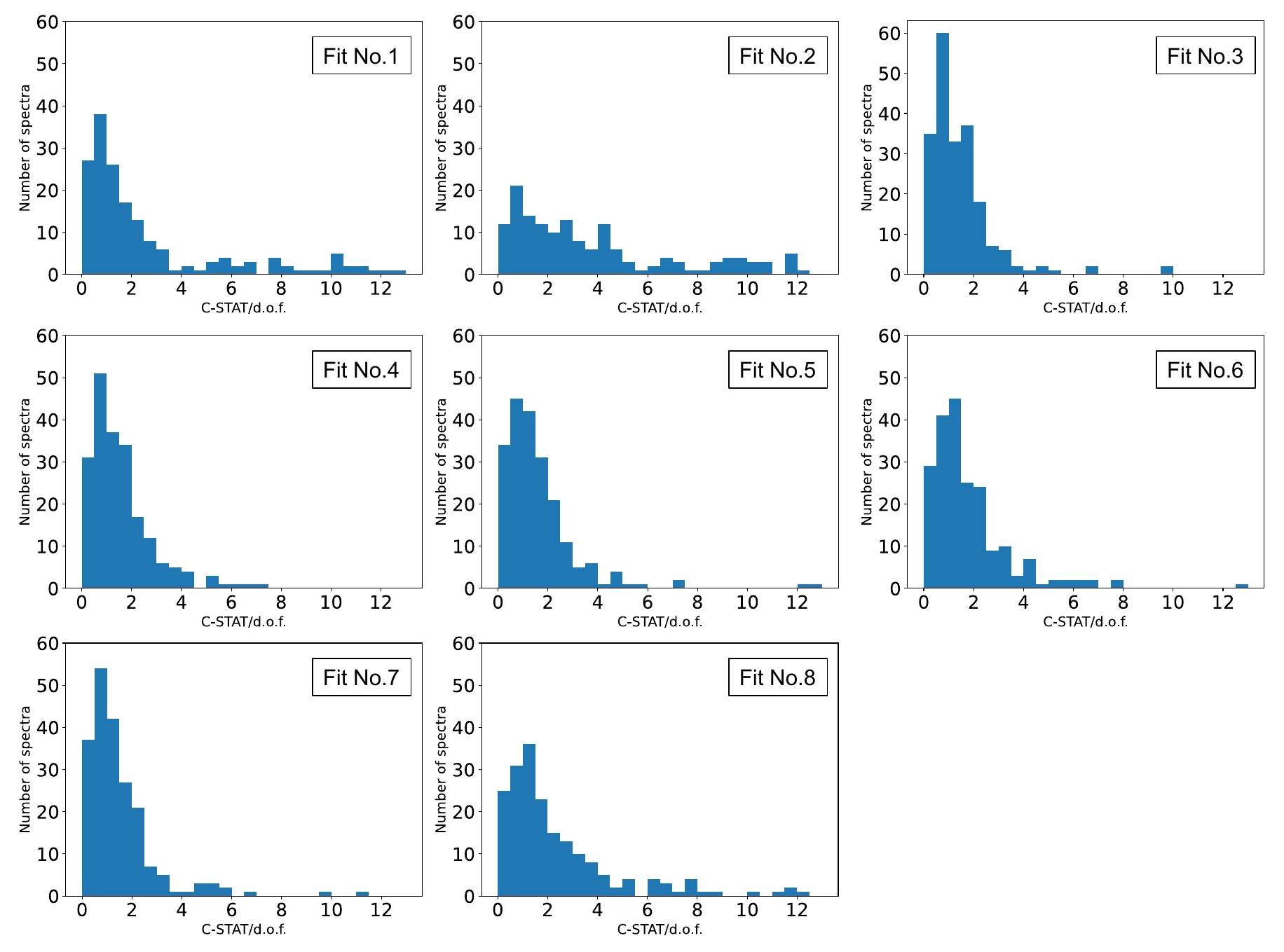}
    \caption{Histograms of the C-stat/d.o.f.~produced by the fits performed in this paper. Labels show from which fit each histogram is derived, and details of each fit are listed in Table~\ref{tb:models}. The third figure is identical to Fig.~\ref{fig:4a}, and other histograms are comparable to it.}
    \label{fig:histall}
\end{figure*}


\newpage
\bsp	
\label{lastpage}
\end{document}